\input harvmac
\input epsf
\input amssym
\baselineskip 13pt

\newcount\figno
\figno=0
\def\fig#1#2#3{
\par\begingroup\parindent=0pt\leftskip=1cm\rightskip=1cm\parindent=0pt
\baselineskip=11pt
\global\advance\figno by 1
\midinsert
\epsfxsize=#3
\centerline{\epsfbox{#2}}
\vskip 12pt
{\bf Fig.\ \the\figno: } #1\par
\endinsert\endgroup\par
}
\def\figlabel#1{\xdef#1{\the\figno}}
\def\encadremath#1{\vbox{\hrule\hbox{\vrule\kern8pt\vbox{\kern8pt
\hbox{$\displaystyle #1$}\kern8pt}
\kern8pt\vrule}\hrule}}

\def\p{\partial}

\def\rt{\rightarrow}
\def\Oc{{\cal O}}

\def\xh{\hat{x}}

\def\av{\vec{a}}

\def\zb{\overline{z}}

\def\xh{\hat{x}}

\def\zb{\overline{z}}

\def\zh{{\hat{{z}}}}

\def\tb{\overline{t}}

\def\ab{\overline{a}}

\def\cb{\overline{c}}

\def\gb{\overline{g}}
\def\fb{\overline{f}}

\def\gt{\tilde{g}}
\def\gh{\hat{g}}

\def\Oc{{\cal O}}

\def\fb{\overline{f}}

\def\gd{\dot{g}}

\def\zd{\dot{z}}

\lref\GiombiVD{
  S.~Giombi, A.~Maloney and X.~Yin,
  ``One-loop Partition Functions of 3D Gravity,''
JHEP {\bf 0808}, 007 (2008).
[arXiv:0804.1773 [hep-th]].
}

\lref\FreedmanTZ{
  D.~Z.~Freedman, S.~D.~Mathur, A.~Matusis and L.~Rastelli,
  ``Correlation functions in the CFT(d) / AdS(d+1) correspondence,''
Nucl.\ Phys.\ B {\bf 546}, 96 (1999).
[hep-th/9804058].
}

\lref\LiuJHS{
  J.~Liu, E.~Perlmutter, V.~Rosenhaus and D.~Simmons-Duffin,
  ``$d$-dimensional SYK, AdS Loops, and $6j$ Symbols,''
JHEP {\bf 1903}, 052 (2019).
[arXiv:1808.00612 [hep-th]].
}

\lref\DolanWY{
  F.~A.~Dolan,
  ``Character formulae and partition functions in higher dimensional conformal field theory,''
J.\ Math.\ Phys.\  {\bf 47}, 062303 (2006).
[hep-th/0508031].
}

\lref\GopakumarQS{
  R.~Gopakumar, R.~K.~Gupta and S.~Lal,
  ``The Heat Kernel on $AdS$,''
JHEP {\bf 1111}, 010 (2011).
[arXiv:1103.3627 [hep-th]].
}

\lref\PenedonesUE{
  J.~Penedones,
  ``Writing CFT correlation functions as AdS scattering amplitudes,''
JHEP {\bf 1103}, 025 (2011).
[arXiv:1011.1485 [hep-th]].
}

\lref\PenedonesNS{
  J.~Penedones,
  ``High Energy Scattering in the AdS/CFT Correspondence,''
[arXiv:0712.0802 [hep-th]].
}

\lref\KrausEZW{
  P.~Kraus, A.~Maloney, H.~Maxfield, G.~S.~Ng and J.~q.~Wu,
  ``Witten Diagrams for Torus Conformal Blocks,''
JHEP {\bf 1709}, 149 (2017).
[arXiv:1706.00047 [hep-th]].
}

\lref\GobeilFZY{
  Y.~Gobeil, A.~Maloney, G.~S.~Ng and J.~q.~Wu,
  ``Thermal Conformal Blocks,''
SciPost Phys.\  {\bf 7}, 015 (2019).
[arXiv:1802.10537 [hep-th]].
}

\lref\FitzpatrickHU{
  A.~L.~Fitzpatrick and J.~Kaplan,
  ``Analyticity and the Holographic S-Matrix,''
JHEP {\bf 1210}, 127 (2012).
[arXiv:1111.6972 [hep-th]].
}

\lref\FitzpatrickZM{
  A.~L.~Fitzpatrick, E.~Katz, D.~Poland and D.~Simmons-Duffin,
  ``Effective Conformal Theory and the Flat-Space Limit of AdS,''
JHEP {\bf 1107}, 023 (2011).
[arXiv:1007.2412 [hep-th]].
}

\lref\DusedauUE{
  D.~W.~Dusedau and D.~Z.~Freedman,
  ``Lehmann Spectral Representation for Anti-de Sitter Quantum Field Theory,''
Phys.\ Rev.\ D {\bf 33}, 389 (1986).
}

\lref\BrosVH{
  J.~Bros, H.~Epstein, M.~Gaudin, U.~Moschella and V.~Pasquier,
  ``Anti de Sitter quantum field theory and a new class of hypergeometric identities,''
Commun.\ Math.\ Phys.\  {\bf 309}, 255 (2012).
[arXiv:1107.5161 [hep-th]].
}

\lref\grig{ A. Grigoryan, M. Noguchi, ``The heat kernel on hyperbolic space", {\it Bull. London Math. Soc.} , {\bf 30} (1998) 643-650}

\lref\SonnerRMT{
  J.~Sonner and B.~Withers,
  ``Linear gravity from conformal symmetry,''
[arXiv:1810.12923 [hep-th]].
}

\lref\CostaKFA{
  M.~S.~Costa, V.~Gonçalves and J.~Penedones,
  ``Spinning AdS Propagators,''
JHEP {\bf 1409}, 064 (2014).
[arXiv:1404.5625 [hep-th]].
}

\lref\AllenWD{
  B.~Allen and T.~Jacobson,
  ``Vector Two Point Functions in Maximally Symmetric Spaces,''
Commun.\ Math.\ Phys.\  {\bf 103}, 669 (1986).
}

\lref\BekaertTVA{
  X.~Bekaert, J.~Erdmenger, D.~Ponomarev and C.~Sleight,
  ``Quartic AdS Interactions in Higher-Spin Gravity from Conformal Field Theory,''
JHEP {\bf 1511}, 149 (2015).
[arXiv:1508.04292 [hep-th]].
}

\lref\MikhailovBP{
  A.~Mikhailov,
  ``Notes on higher spin symmetries,''
[hep-th/0201019].
}

\lref\BekaertTVA{
  X.~Bekaert, J.~Erdmenger, D.~Ponomarev and C.~Sleight,
  ``Quartic AdS Interactions in Higher-Spin Gravity from Conformal Field Theory,''
JHEP {\bf 1511}, 149 (2015).
[arXiv:1508.04292 [hep-th]].
}

\lref\HeemskerkPN{
  I.~Heemskerk, J.~Penedones, J.~Polchinski and J.~Sully,
  ``Holography from Conformal Field Theory,''
JHEP {\bf 0910}, 079 (2009).
[arXiv:0907.0151 [hep-th]].
}

\lref\FitzpatrickDM{
  A.~L.~Fitzpatrick and J.~Kaplan,
  ``Unitarity and the Holographic S-Matrix,''
JHEP {\bf 1210}, 032 (2012).
[arXiv:1112.4845 [hep-th]].
}

\lref\FitzpatrickYX{
  A.~L.~Fitzpatrick, J.~Kaplan, D.~Poland and D.~Simmons-Duffin,
  ``The Analytic Bootstrap and AdS Superhorizon Locality,''
JHEP {\bf 1312}, 004 (2013).
[arXiv:1212.3616 [hep-th]].
}

\lref\CaronHuot{
S.~Caron-Huot,
``Analyticity in Spin in Conformal Theories,''
JHEP {\bf {09}} (2017), 078
doi:10.1007/JHEP09(2017)078
[arXiv:1703.00278 [hep-th]].}

\lref\Fitzpatrickhh{
A.~Fitzpatrick and D.~Shih,
``Anomalous Dimensions of Non-Chiral Operators from AdS/CFT,''
JHEP  {bf{10}} (2011), 113
doi:10.1007/JHEP10(2011)113
[arXiv:1104.5013 [hep-th]].}


\lref\Aldayvkk{
L.~F.~Alday and S.~Caron-Huot,
``Gravitational S-matrix from CFT dispersion relations,''
JHEP {\bf 12} (2018), 017
doi:10.1007/JHEP12(2018)017
[arXiv:1711.02031 [hep-th]].}

\lref\Carmiqzm{
D.~Carmi, L.~Di Pietro and S.~Komatsu,
``A Study of Quantum Field Theories in AdS at Finite Coupling,''
JHEP  {\bf 01} (2019), 200
doi:10.1007/JHEP01(2019)200
[arXiv:1810.04185 [hep-th]].}

\lref\Liujhs{
J.~Liu, E.~Perlmutter, V.~Rosenhaus and D.~Simmons-Duffin,
``$d$-dimensional SYK, AdS Loops, and $6j$ Symbols,''
JHEP {\bf 03} (2019), 052
doi:10.1007/JHEP03(2019)052
[arXiv:1808.00612 [hep-th]].}

\lref\Cardonadov{
C.~Cardona and K.~Sen,
``Anomalous dimensions at finite conformal spin from OPE inversion,''
JHEP {\bf 11} (2018), 052
doi:10.1007/JHEP11(2018)052
[arXiv:1806.10919 [hep-th]].}

\lref\Sleightryu{
C.~Sleight and M.~Taronna,
``Anomalous Dimensions from Crossing Kernels,''
JHEP {\bf 11} (2018), 089
doi:10.1007/JHEP11(2018)089
[arXiv:1807.05941 [hep-th]].}


\lref\Liuth{
H.~Liu,
``Scattering in anti-de Sitter space and operator product expansion,''
Phys.\ Rev.\ D {\bf 60}, 106005 (1999)
doi:10.1103/PhysRevD.60.106005
[arXiv:hep-th/9811152 [hep-th]].}

\lref\DHokerecp{
E.~D'Hoker and D.~Z.~Freedman,
``General scalar exchange in AdS(d+1),''
Nucl.\ Phys.\ B {\bf 550}, 261-288 (1999)
doi:10.1016/S0550-3213(99)00169-8
[arXiv:hep-th/9811257 [hep-th]].}

\lref\DHokerkzh{
E.~D'Hoker, D.~Z.~Freedman, S.~D.~Mathur, A.~Matusis and L.~Rastelli,
``Graviton exchange and complete four point functions in the AdS / CFT correspondence,''
Nucl.\ Phys.\ B {\bf 562}, 353-394 (1999)
doi:10.1016/S0550-3213(99)00525-8
[arXiv:hep-th/9903196 [hep-th]].}

\lref\DHokermic{
E.~D'Hoker, S.~D.~Mathur, A.~Matusis and L.~Rastelli,
``The Operator product expansion of N=4 SYM and the 4 point functions of supergravity,''
Nucl.\ Phys.\ B {\bf 589}, 38-74 (2000)
doi:10.1016/S0550-3213(00)00523-X
[arXiv:hep-th/9911222 [hep-th]].}

\lref\Hoffmannmx{
L.~Hoffmann, A.~C.~Petkou and W.~Ruhl,
``Aspects of the conformal operator product expansion in AdS / CFT correspondence,''
Adv.\ Theor.\ Math.\ Phys.\  {\bf 4}, 571-615 (2002)
doi:10.4310/ATMP.2000.v4.n3.a3
[arXiv:hep-th/0002154 [hep-th]].}

\lref\Arutyunovku{
G.~Arutyunov, S.~Frolov and A.~C.~Petkou,
``Operator product expansion of the lowest weight CPOs in $ N=4$ SYM$_4$ at strong coupling,''
Nucl.\ Phys.\ B {\bf 586}, 547-588 (2000)
doi:10.1016/S0550-3213(00)00439-9
[arXiv:hep-th/0005182 [hep-th]].}

\lref\Hoffmanntb{
L.~Hoffmann, L.~Mesref and W.~Ruhl,
``AdS box graphs, unitarity and operator product expansions,''
Nucl.\ Phys.\ B {\bf 589}, 337-355 (2000)
doi:10.1016/S0550-3213(00)00517-4
[arXiv:hep-th/0006165 [hep-th]].}

\lref\Arutyunovrs{
G.~Arutyunov, S.~Penati, A.~Petkou, A.~Santambrogio and E.~Sokatchev,
``Nonprotected operators in N=4 SYM and multiparticle states of AdS(5) SUGRA,''
Nucl.\ Phys.\ B {\bf 643}, 49-78 (2002)
doi:10.1016/S0550-3213(02)00679-X
[arXiv:hep-th/0206020 [hep-th]].}

\lref\Cornalbaxm{
L.~Cornalba, M.~S.~Costa, J.~Penedones and R.~Schiappa,
``Eikonal Approximation in AdS/CFT: Conformal Partial Waves and Finite N Four-Point Functions,''
Nucl.\ Phys.\ B {\bf 767}, 327-351 (2007)
doi:10.1016/j.nuclphysb.2007.01.007
[arXiv:hep-th/0611123 [hep-th]].}

\lref\Cornalbazb{
L.~Cornalba, M.~S.~Costa and J.~Penedones,
``Eikonal approximation in AdS/CFT: Resumming the gravitational loop expansion,''
JHEP {\bf 09}, 037 (2007)
doi:10.1088/1126-6708/2007/09/037
[arXiv:0707.0120 [hep-th]].}

\lref\Heemskerkty{
I.~Heemskerk and J.~Sully,
``More Holography from Conformal Field Theory,''
JHEP {\bf 09}, 099 (2010)
doi:10.1007/JHEP09(2010)099
[arXiv:1006.0976 [hep-th]].}


\lref\Hellermanbu{
S.~Hellerman,
``A Universal Inequality for CFT and Quantum Gravity,''
JHEP { \bf 08}, 130 (2011)
doi:10.1007/JHEP08(2011)130

[arXiv:0902.2790 [hep-th]].}

\lref\Hijanozsa{
E.~Hijano, P.~Kraus, E.~Perlmutter and R.~Snively,
``Witten Diagrams Revisited: The AdS Geometry of Conformal Blocks,''
 JHEP  {\bf 01}, 146 (2016)
[arXiv:1508.00501 [hep-th]].}

\lref\BrosVH{
  J.~Bros, H.~Epstein, M.~Gaudin, U.~Moschella and V.~Pasquier,
  ``Anti de Sitter quantum field theory and a new class of hypergeometric identities,''
Commun.\ Math.\ Phys.\  {\bf 309}, 255 (2012).
[arXiv:1107.5161 [hep-th]].
}


\lref\HartmanDY{
  T.~Hartman and L.~Rastelli,
  ``Double-trace deformations, mixed boundary conditions and functional determinants in AdS/CFT,''
JHEP {\bf 0801}, 019 (2008).
[hep-th/0602106].
}

\lref\DiazAN{
  D.~E.~Diaz and H.~Dorn,
  ``Partition functions and double-trace deformations in AdS/CFT,''
JHEP {\bf 0705}, 046 (2007).
[hep-th/0702163 [HEP-TH]].
}

\lref\DenefKN{
  F.~Denef, S.~A.~Hartnoll and S.~Sachdev,
  ``Black hole determinants and quasinormal modes,''
Class.\ Quant.\ Grav.\  {\bf 27}, 125001 (2010).
[arXiv:0908.2657 [hep-th]].
}

\lref\GopakumarQS{
  R.~Gopakumar, R.~K.~Gupta and S.~Lal,
  ``The Heat Kernel on $AdS$,''
JHEP {\bf 1111}, 010 (2011).
[arXiv:1103.3627 [hep-th]].
}

\lref\DHokerBVE{
  E.~D'Hoker, D.~Z.~Freedman, S.~D.~Mathur, A.~Matusis and L.~Rastelli,
  ``Graviton and gauge boson propagators in AdS(d+1),''
Nucl.\ Phys.\ B {\bf 562}, 330 (1999).
[hep-th/9902042].
}

\lref\NaqviVA{
  A.~Naqvi,
  ``Propagators for massive symmetric tensor and p forms in AdS(d+1),''
JHEP {\bf 9912}, 025 (1999).
[hep-th/9911182].
}

\lref\SimmonsDuffinNUB{
  D.~Simmons-Duffin, D.~Stanford and E.~Witten,
  ``A spacetime derivation of the Lorentzian OPE inversion formula,''
JHEP {\bf 1807}, 085 (2018).
[arXiv:1711.03816 [hep-th]].
}

\Title{\vbox{\baselineskip14pt
}} {\vbox{\centerline { Anomalous Dimensions  from}\vskip.3cm  \centerline {  Thermal AdS Partition Functions}}}
\centerline{Per Kraus$^{1\dagger}$, Stathis Megas$^{1 \ddagger}$, Allic Sivaramakrishnan$^{2*}$}
\bigskip
\centerline{\it{$^1$Mani L. Bhaumik Institute for Theoretical Physics}}
\centerline{${}$\it{University of California, Los Angeles, CA 90095, USA}}
\centerline{\it{$^2$Department of Physics and Astronomy}}
\centerline{${}$\it{University of Kentucky, Lexington, KY 40506, USA}}

\baselineskip14pt

\vskip .3in

\centerline{\bf Abstract}
\vskip.2cm

\noindent

We develop an efficient method for computing thermal partition functions of weakly coupled scalar fields in AdS. We consider quartic contact interactions and show how to evaluate the relevant two-loop vacuum diagrams without performing any explicit AdS integration, the key step being the use of K{\"a}ll{\'e}n-Lehmann type identities. This leads to a simple method for extracting double-trace anomalous dimensions in any spacetime dimension, recovering known first-order results in a streamlined fashion.

\vskip 2.1in
\noindent
------------------------------------------------
\vskip -2.1in
\Date{\tt{${^\dagger}$pkraus@ucla.edu, ${^\ddagger}$stathismegas@physics.ucla.edu, ${^*}$allicsiva@uky.edu}}

\baselineskip14pt

\listtoc \writetoc

\newsec{Introduction}

It is interesting both for its own sake and in connection to the AdS$_{d+1}$/CFT$_d$ correspondence to ask how the spectrum of a weakly coupled quantum field theory in AdS behaves as a function of its coupling. On the AdS side, this amounts to computing binding energies of multi-particle states \refs{\Liuth\DHokerecp\DHokerkzh\DHokermic\Hoffmannmx
\Arutyunovku\Hoffmanntb\Arutyunovrs\Cornalbaxm\Cornalbazb}-\Heemskerkty, while on the CFT side it corresponds to computing anomalous dimensions of multi-trace operators.   Much effort has gone into such computations in the context of the bootstrap program following \HeemskerkPN, in which the anomalous dimensions, along with the OPE coefficients, comprise the CFT data.  The original approach to extracting anomalous dimensions is to expand a correlation function in conformal blocks.  While straightforward in principle, in practice the details can be rather messy, particularly for odd $d$, where the conformal blocks do not have closed form expressions.   Notably though, a major simplification for handling tree level exchange diagrams is provided by the Lorentzian inversion formula \CaronHuot, which bypasses the need to compute the full Witten diagram \CaronHuot\Aldayvkk\Cardonadov\Sleightryu\Liujhs\SimmonsDuffinNUB .  Furthermore, as we  discuss in  Appendix C, for the type of interactions considered in this paper powerful harmonic analysis techniques are available that do not require the explicit conformal blocks.  Another approach offering some simplifications is to compute energy shifts using standard quantum mechanical perturbation theory \FitzpatrickZM,\Fitzpatrickhh.

In this paper we develop a different approach: we compute thermal partition functions and extract anomalous dimensions by expanding bulk vacuum diagrams in characters.\foot{Constraints on the AdS$_3$/CFT$_2$ spectrum implied by modular invariance of the partition function form the basis of the modular bootstrap program \Hellermanbu.  Modular invariance (for $d=2$) will play no role for us, since we focus on the low energy spectrum without regard to issues of UV completion.}   We focus on quartic contact diagrams, with various numbers of derivatives.  Given that our method is designed to extract anomalous dimensions but not OPE coefficients, one might expect that it involves less work than a correlation function based approach, and we indeed find this to be the case.  It is easy to work out results in arbitrary spacetime dimension, as we illustrate with various worked examples.   A key simplification is that the conformal characters have simple graphical AdS representations, allowing one to expand the partition function in characters without having to perform any integrals.  This simplification is similar to the one provided by the use of geodesic Witten diagrams \Hijanozsa.

Since the main elements in our approach, and their implementation, are simple to explain, in the remainder of this section we describe all the steps involved in extracting the anomalous dimensions for the basic $\lambda \phi^4$ interaction, and also indicate how to incorporate derivative interactions, with full details provided in the main body of the text.

\subsec{General method and summary of results}

Thermal AdS$_{d+1}$ is described by the Euclidean signature line element
\eqn\ak{ ds^2 = {1\over \cos^2 \rho}\left(d\rho^2 +dt^2 +\sin^2 \rho d\Omega_{d-1}^2 \right)~,}
with periodic imaginary time,
\eqn\ab{ t \cong t+\beta~. }
Here and elsewhere we are setting the AdS radius to unity.
We consider some weakly interacting quantum field theory with coupling constant $\lambda$ living on this background geometry, and seek to compute the thermal partition sum
\eqn\ac{ Z(\beta) = \Tr e^{-\beta H}~,}
where $H$ denotes the Hamiltonian generating translations of $t$.\foot{A more general partition sum $Z(\beta,\mu_a)$  would include chemical potentials $\mu_a$ conjugate to the Cartan generators of the SO(d) rotation group, as in Appendix A.   We mainly focus on $Z(\beta)$.}   The form of $Z(\beta)$ is dictated by the isometry group of AdS$_{d+1}$.   The spectrum can be organized into unitary irreducible representations of the Lorentzian isometry group.     Each such representation is labelled by a scaling dimension\foot{so-called because $H$ can alternatively be realized as the dilatation generator acting on $R^d$.}  $\Delta$ and representation $R$ of the SO(d) rotation group acting on the angular coordinates in \ak.  The lowest energy (primary) states in each representation obey $H|\Delta;R\rangle = \Delta |\Delta;R\rangle$, and excited states are obtained by acting with generators $P_a$ ($a=1,2,\ldots d$), each of which raises the eigenvalue of $H$ by one unit.\foot{Certain representations, as arise in the case of gauge fields,   have null states and so this statement requires modification, but this will not be relevant to our considerations.}   The character of a given representation labelled by $(\Delta, R)$ is then given by
\eqn\ad{\chi_{\Delta,R}(\beta) = \Tr_{\Delta,R} e^{-\beta H} = {d_R q^\Delta \over (1-q)^d}~,}
where $d_R$ is the dimension of the SO(d) representation $R$, and
\eqn\ae{ q=e^{-\beta}~.}
The general partition sum may be expressed as a sum of characters,
\eqn\af{ Z(\beta) = 1+\sum_{\Delta,R}  N_{\Delta,R} \chi_{\Delta,R}(\beta)~,}
where $ N_{\Delta,R}$ denotes the multiplicity.

In the free theory at $\lambda=0$ the Hilbert space is a Fock space of single and multi-particles states.  For example, consider a free scalar field of mass $m$.  As is well known, the single particle primary is an SO(d) singlet and carries energy $\Delta$ related to $m^2$ by the equation $m^2 =\Delta(\Delta-d)$.   Two-particle primaries are described by bound states with radial quantum number $n=0,1,2,\ldots$ and angular momentum quantum number $J$.  $J$ denotes  a symmetric traceless tensor representation of SO(d).   Due to Bose symmetry, $J$ takes only even (non-negative) integer values, $J=0,2,4,\ldots$.    The scaling dimensions are $2\Delta+2n +J$.     The partition sum of such a free scalar is therefore
\eqn\ag{ Z(\beta) = 1 + \chi_{\Delta,0} + \sum_{J=0,2,\ldots} \sum_{n=0}^\infty \chi_{2\Delta+2n +J,J}(\beta) + \ldots~,}
where $\ldots$ denotes the contribution from states with three or more particles.   In the AdS$_{d+1}$/CFT$_d$ correspondence, the CFT states dual to the one and two particles states in the bulk are created by ``single-trace" and ``double-trace" operators, whose representation labels match the above $(\Delta, R)$ assignments.  Schematically, $[\Oc\Oc]_{n,J} \sim \Oc \p_{\mu_1} \ldots \p_{\mu_J} (\p^2)^n \Oc$.

We now turn on the coupling constant $\lambda$, which we take to vanish in the large-$N$ limit.  This preserves the symmetry group and so the partition function can still be expanded in characters as in \af.  The SO(d) representations cannot change continuously, but the scaling dimensions $\Delta$ can.  What is meaningful is the relation between the energies of the multi-particle states and the single particle states.  We think of keeping the single particle dimension $\Delta$ fixed as we turn on $\lambda$; alternatively, we trade the two parameters $(m^2,\lambda)$ for $(\Delta_{1-{\rm particle}},\lambda)$.   The partition sum \ag\ is then modified by the replacement
\eqn\ah{ 2\Delta+2n +J \rt 2\Delta+2n +J + \gamma(n,J).}
The  energy shift $\gamma(n,J)$ is interpreted in the dual CFT as an anomalous dimension acquired by the double trace operator $[\Oc\Oc]_{n,J}$.
More generally, to first order in $\lambda $ we can write
\eqn\ai{ Z_\lambda(\beta) = Z_0(\beta) + \sum_{\Delta,R}  N_{\Delta,R} {\p \chi_{\Delta,R}(\beta)\over \p \Delta} \gamma(n,J)~.}
If we can set up the computation of $Z_\lambda(\beta)$ such that it naturally takes the form \ai, then we can easily read off the corresponding anomalous dimensions.   We now describe the strategy for doing so.

We will compute the partition function from the functional integral in thermal AdS$_{d+1}$.  A simple relation of central importance here is the bulk representation of the character $\chi_{\Delta,J}(\beta)$.   These characters correspond to diagrams in which a spin-$J$ particle winds  once around the thermal circle, the precise relation being
\eqn\aj{ \chi_{\Delta_J,J}(\beta) =  \int_{m_J^2}^\infty\! d{m'}_J^2 \int\! d^{d+1} x\sqrt{g} \Tr \left[ \Pi_{\Delta'_J,J} (x,x_{\beta})\right]~. }
Here $\Pi_{\Delta_J,J}(x,y)$  (indices suppressed) is the spin-$J$ bulk-bulk propagator;\foot{Note that in the spin-$0$ case we often write $\Pi_{\Delta,0}(x,y)=G_\Delta(x,y)$.}  $x_\beta=(t+\beta,\rho,\Omega)$ denotes the bulk coordinate $x$ displaced by one thermal translation, and the relation between the mass and scaling dimension is
\eqn\ak{ m_J^2= \Delta_J(\Delta_J-d)-J~.}
The integral in \aj\ is over thermal AdS.      We therefore have
\eqn\al{ {\p  \chi_{\Delta_J,J}(\beta) \over \p \Delta_J} = -(2\Delta_J-d)  \int\! d^{d+1} x\sqrt{g} \Tr \left[ \Pi_{\Delta_J,J} (x,x_{\beta})\right]~. }
To summarize our method: to find anomalous dimensions, we expand bulk diagrams in the object above and then read off the coefficients. Being the bulk dual of ${\p  \chi_{\Delta_J,J}(\beta) \over \p \Delta_J}$, the integrated propagator here is analogous to the geodesic Witten diagram in the context of correlation functions \Hijanozsa. This new way of computing anomalous dimensions therefore has similar advantages to geodesic Witten diagram methods.

In this work we focus on scalar fields interacting via quartic contact interactions with any number of derivatives.   To explain the basic idea for treating these, we first consider the simplest case of a single scalar field $\phi$ with $S_{\rm int} = \lambda \int d^{d+1}x\sqrt{g}\phi^4$.  At order $\lambda$ the partition function, obtained by Wick contraction, is
\eqn\am{ \ln Z_\lambda(\beta) = \ln Z_0(\beta) -3\lambda \int d^{d+1}x\sqrt{g} G_\Delta^\beta(x,x) G_\Delta^\beta(x,x)~.}
Here $G_\Delta^\beta(x,x)$ is the  scalar thermal bulk-bulk propagator, which can be constructed from the global propagator by summing over thermal  images, $G_\Delta^\beta(x,x) = \sum_{n=-\infty}^\infty G_\Delta(x,x_{n\beta})$, with $x_{n\beta}=(t+n\beta,\rho,\Omega)$. We focus on two-particle states, since effects on states with more than two particles requires going to higher order in $\lambda$.  Since the $n$th term in the sum corresponds to a particle winding $n$ times around the thermal circle, and hence to an $n$ particle state if viewed at an instant of time, for two-particles states we  keep only the  $n=\pm 1$ terms from each image sum. Each gives the same contribution, and so we have 
\eqn\am{ \ln Z_\lambda(\beta) = \ln Z_0(\beta) -12\lambda \int d^{d+1}x\sqrt{g} G_\Delta(x,x_\beta) G_\Delta(x,x_\beta)~ +\ldots}
the $\ldots$ now denoting contributions from states of three or more particles.

We now make the link to \ai\ by using the identity
\eqn\an{  G_\Delta(x,y) G_\Delta(x,y) = \sum_{n=0}^\infty a_n^{(0)} G_{2\Delta+2n}(x,y)~,}
with coefficients $a_n^{(0)}$  given below in 1.23.  The existence of such an identity, which can be thought of as a version of a K{\"a}ll{\'e}n-Lehmann representation \DusedauUE\FitzpatrickHU\FitzpatrickZM\ (we note that the latter two references use this identity in a context similar to ours), is readily understood by comparing terms in a series expansion in the inverse geodesic distance\foot{The existence of this identity can also be understood from properties of harmonic functions, for which we refer the reader to \PenedonesUE,\CostaKFA.}.   Using this identity, together with \al, we have
\eqn\ao{  \ln Z_\lambda(\beta) = \ln Z_0(\beta) +6\lambda \sum_{n=0}^\infty {a_n^{(0)}\over 2\Delta+2n-d/2} {\p \chi_{\Delta,0}(\beta)\over \p \Delta}\Big|_{2\Delta+2n} ~ +\ldots}
From \ai\ we then read off the anomalous dimensions to first order in $\lambda$ as
\eqn\ap{ \gamma(n,0) = {6a_n^{(0)} \lambda  \over 2\Delta+2n-d/2}~,}
and  with $\gamma(n,J)=0$ for $J>0$.  An especially simple case is AdS$_3$  ($d=2$) for which $G_\Delta(x,y) = {1\over 2\pi} {e^{-\Delta \sigma(x,y)}/(1-e^{-2\sigma(x,y)})}$, where $\sigma(x,y)$ is the geodesic distance.  It is then a triviality to see that $a_n^{(0)} = {1\over 2\pi}$ satisfies \an.  This illustrates how relatively  little work is required in this approach.  Using  1.23, with $s=0$ and $\Delta_1=\Delta_2=\Delta$, gives the spin-$0$ anomalous dimensions for general $d$, in agreement with the known result \FitzpatrickZM.  The corresponding result for a pair of scalar fields with interaction $\lambda \int\! (\phi_1 \phi_2)^2$ follows from \ap\ once we divide by $3$ (due to the single Wick contraction) and replace $2\Delta \rt \Delta_1+\Delta_2$.

For vertices with derivatives we will need a generalized version of the identity \an.   We write the scalar propagator as $G_\Delta(u)$, where $u=u(x,y)$ is the (half) chordal-distance, related to the geodesic distance as
\eqn\aq{  u(x,y) = -1+ \cosh \sigma(x,y)~.}
In terms of this variable the scalar propagator is
\eqn\ar{  G_\Delta(u) = C_\Delta \big(2u)^{-\Delta}  F\left(\Delta,\Delta-{d-1\over 2},2\Delta-d+1;-{2\over u} \right)}
with
\eqn\as{ C_\Delta = {\Gamma(\Delta) \over 2\pi^{d/2}\Gamma(\Delta+1-{d\over 2}) }   }
and where $F$ denotes the ${_2}F_1$ hypergeometric function.   The general identity we need is
\eqn\at{ \eqalign{   G^{(s)}_{\Delta_1}(u)  G^{(s)}_{\Delta_2}(u)  & =   \sum_{n=0}^\infty a^{(s)}_n   G^{(s)}_{\Delta_1+\Delta_2+2n+s}(u)  }}
where we are using the notation
\eqn\au{  G^{(s)}_{\Delta}(u) = {d^s \over du^s} G_{\Delta}(u)~.}
Again, the existence of the relation \at\ follows from comparing the expansions in $1/u$, and the coefficients are found to be
\eqn\av{\eqalign{  a^{(s)}_n  &= { (-2)^s (h+s)_n \over 2\pi^h n!}{(\Delta_1+\Delta_2+2n+2s)_{1-h-s} (\Delta_1+\Delta_2-2h+n+1)_{n}\over (\Delta_1+n+s)_{1-h-s}(\Delta_2+n+s)_{1-h-s} (\Delta_1+\Delta_2-h+n+s)_n}~,  }}
with $ h \equiv {d/ 2}$ and the Pochhammer symbol is $(a)_n \equiv {\Gamma(a+n)\over \Gamma(a)}$.
For example, consider the interaction $S_{\rm int} = \lambda\int\! d^{d+1}x\sqrt{g} \phi^2 (\nabla_{\mu_1} \ldots \nabla_{\mu_J} \phi)^2$ with $J=2,4,6, \ldots$.  Such a vertex is known to give rise to anomalous dimensions for double-trace operators of spin $s=0,2,\ldots J$.   As we'll see, the highest spin contribution is very easy to extract using our approach.  We need to expand the product of two differentiated scalar propagators in terms of spinning propagators and their derivatives.  The spinning propagators can be expressed in terms of the scalar propagator, which will lead us to the identity \at. The result for the spin-$J$ anomalous dimensions is simply
\eqn\ay{ \gamma(n,J) ={8J! a_n^{(J)}  \lambda \over 2\Delta +2n+J -h}~.}
This result matches that in \HeemskerkPN, which was obtained by decomposing a four-point function in the Regge limit.     The anomalous dimension for spins $s<J$ are also straightforward to extract in principle, although the details require more bookkeeping.   We work out the full details in the case of two and four derivative interactions, the former requiring two distinct fields in order to be nontrivial (i.e. not reducible to $\phi^4$ after using integration by parts and the free field equations).  The four derivative example gives results for general $d$ which reproduce a known  expression for $d=2$.  Another  interaction which is easily handled is $S_{\rm int} = \lambda \int\! d^{d+1}x\sqrt{g}   \nabla^\mu \phi\nabla_\mu \phi (\nabla_{\mu_1} \ldots \nabla_{\mu_J} \phi)^2$ with $J=2,4,6, \ldots$.  This $2J+2$ derivative interaction gives rise to anomalous dimensions of operator up to spin-$J$ and, as before, it is simple to extract the highest spin result.   The result for $\gamma(n,J)$ is given by \ay\ multiplied by a factor $m_J^2(\Delta_{n,J})$, where $\Delta_{n ,J}= 2\Delta+2n+J$.  In summary, compared to previous approaches, the  partition function based approach proves to be efficient and involves relatively elementary ingredients.

 The remainder of this paper is organized as follows.  In section 2 we recall some basic facts about CFT characters and partition functions.  In section 3 we show how to derive free field partition functions in AdS from the path integral.  This computation is not strictly needed for the rest of the paper, but we have included it since it is a basic result, and one that we have not seen presented in general spacetime dimension.    In section 4 we show how to compute anomalous dimensions using our approach applied to several choices of contact interaction.   In section 5 we review needed facts about AdS propagators for massive symmetric tensor fields.  In section 6 we discuss the general procedure for studying an arbitrary quartic contact interaction, and then apply this to another example in section 7.  Some discussion appears in section 8.  In Appendix A we show how to include angular potentials into our free field partition function computation.  Appendix B sketches the computation of the free partition function of a massive spin-1 field.   Appendix C discusses an alternative approach to extracting anomalous dimensions, based on applying harmonic analysis techniques to the boundary four-point function.

\newsec{CFT partition functions}

We consider CFT$_{d}$ defined on $S^{d-1}\times R$.  The basic partition function is
\eqn\ba{ Z(\beta) = \Tr e^{-\beta H} }
where $H$ is the Hamiltonian generating time translations.  More generally, we can consider $Z(\beta;\mu_i)$, where $\mu_i$ are angular chemical potentials conjugate to the Cartan generators of SO(d).  The Hilbert space is described via the state-operator map, and can be decomposed into irreducible representations of the Lorentzian conformal group $SO(d,2)$.  Each representation is specified by a primary operator $\Oc_{\Delta, R}$, where $R$ denotes a SO(d) representation, corresponds to a state (or rather an SO(d) multiplet of states)  of energy $\Delta$,  $H|\Delta; R\rangle = \Delta |\Delta; R\rangle$.
%
%
%
%
The characters were written down in \ad, and the partition function is given as a sum of characters as in \af.

In a large $N$ CFT, dual to a weakly coupled theory in AdS, the spectrum of primary operators can be organized into single trace primaries and their multi-trace products.  This corresponds in the bulk to a description of the Hilbert space in terms of single-particle and multi-particle states. Consider a spinless single trace primary operator $\Oc_\Delta$.  In the large $N$ limit there exist double trace  primary operators of schematic form $[\Oc \Oc]_{2\Delta+2n +J,J} \sim \Oc \p_{\mu_1} \ldots \p_{\mu_J} (\p^2)^{n} \Oc$.  These operators transform in the rank-$J$ symmetric, traceless tensor representation of SO(d), and Bose symmetry requires $J$ to be an even integer.  There similarly exist triple trace, quadruple trace, $\ldots$,  primary operators.  The contribution to the partition function from such operators is most easily written down by thinking (simply as a mnemonic)  in terms of the dual bulk description.  Quantizing a free scalar field in the bulk yields single particle states with quantum numbers in correspondence with those of the primary operator $\Oc_\Delta$ and its conformal descendants,  $(P_{1})^{m_1}  \ldots  (P_{d})^{m_d} \cdot \Oc_\Delta$.  Focusing on a mode with specified $\{m_a\}$ the contribution to the partition function corresponds to summing over the occupation number $k$,
\eqn\bd{ Z_{\{m_a\}} = \sum_{k=0}^\infty q^{(\Delta + \sum_a m_a)k }={1\over 1- q^{\Delta + \sum_a m_a}}.
}
The total contribution to the partition from the scalar field is then obtained by taking the product over all modes,
\eqn\bez{ Z =\prod_{\{m_a\}}Z_{\{m_a\}} = \prod_{\{m_a\}} {1\over 1- q^{\Delta + \sum_a m_a}}~. }
We rewrite this as
\eqn\bfz{\eqalign{ \ln Z& = -\sum_{\{m_a\}} \ln \left(1- q^{\Delta + \sum_a m_a}  \right)  \cr
& = \sum_{\{m_a\}} \sum_{k=1}^\infty {1\over k} q^{(\Delta + \sum_a m_a)k}\cr
& =  \sum_{k=1}^\infty  {1\over k} {q^{k\Delta} \over (1-q^k)^d}~.    }}
The contribution to $Z$ from $n$ particle states is identified by an overall factor of $q^{n\Delta}$.   For example, the contribution to the partition functions from states of fewer than three particles is
\eqn\bg{ Z =1 +{q^\Delta \over (1-q)^d} + {1\over 2}\left( {1\over (1-q^2)^d }+{1\over (1- q)^{2d} } \right) q^{2\Delta} + \ldots }
We can read off the spectrum of double trace operators by writing the $q^{2\Delta}$  term as
\eqn\bh{ {1\over 2}\left( {1\over (1-q^2)^d }+{1\over (1- q)^{2d} } \right) q^{2\Delta} = \sum_{J=0,2,4,\ldots} \sum_{n=0}^\infty \chi_{2\Delta+2n+J,J}(q)  }
where
\eqn\bha{ \chi_{2\Delta+2n+J,J}(q)=d_J \chi_{2\Delta+2n+J,0}(q)  =  d_J{q^{2\Delta+ 2n+J}~\over (1-q)^d} ,}
is a spin-$J$ character written in terms of the spin-$0$ character
and
\eqn\bi{ d_J = {\Gamma(d+J) \over \Gamma(d)\Gamma(J+1) } -  {\Gamma(d+J-2) \over \Gamma(d)\Gamma(J-1) } }
is the dimension of the rank-$J$ symmetric traceless tensor representation of SO(d).  The right hand side of \bh\ is thus identified with the expected sum over primary operators, one for each even $J$,  with the factor  $1/(1-q)^d$ in \bha\ coming from the sum over descendants.

Now we go to the next order in the $1/N$ expansion.  This corresponds to introducing an interaction term in the bulk with some coupling constant $\lambda$ and working to first order in $\lambda$.  By convention, we continue to write the dimension of the single trace operator  as $\Delta$, absorbing any $\lambda$ dependence into the definition of $\Delta$.   The multi-trace operators pick up anomalous dimensions.  We write the dimension of the  double trace operators as
\eqn\bj{ 2\Delta + 2n + J + \gamma(n,J) }
with $\gamma(n,J)  = O(\lambda)$.   We can therefore write the partition function to first order in $\lambda$ as
\eqn\bk{ \ln Z = \ln Z\big|_{\lambda=0} +  \sum_{J=0,2,4,\ldots} \sum_{n=0}^\infty {\p\chi_{\Delta,J}(q) \over \p\Delta}\Big|_{\Delta=2\Delta+2n+J}\gamma(n,J) + \ldots ~,}
%
where again we are only including single and double trace operators.   This is the expression we will use to read off anomalous dimensions.  We will compute $\ln Z$ to first order in $\lambda$ and write the result in the form \bk, and thereby read off $\gamma(n,J)$.

Alternatively, we can consider a pair of scalar primary operators, $\Oc_1$ and $\Oc_2$.   We form double trace operators as before, $[\Oc_1 \Oc_2]_{\Delta_1+\Delta_2+2n+J,J} \sim \Oc_1 \p_{\mu_1} \ldots \p_{\mu_J} (\p^2)^n \Oc_2$ except that now $J$ runs over all non-negative integers,  since there is no Bose symmetry.   The anomalous dimensions are read off from the expression
\eqn\bl{ \ln Z = \ln Z\big|_{\lambda=0} +  \sum_{J=0,1,2,\ldots} \sum_{n=0}^\infty {\p\chi_{\Delta,J}(q) \over \p\Delta}\Big|_{\Delta=\Delta_1+\Delta_2+2n+J}\gamma(n,J) + \ldots ~}

\newsec{Computation of free field partition functions in AdS$_{d+1}$}

For completeness, in this section we discuss the path integral computation of free field partition functions in thermal AdS.  These computations will in fact not be needed for the main goal of this work, which is the extraction of anomalous dimensions from the interacting theory.  The reason is that the contribution from the interaction vertex will automatically take the form \bk-\bl.   Nevertheless, it is a useful exercise to see how the free scalar field partition function arises from the path integral, which  as far as we know this has not been done in general dimension $d$.   See  \HartmanDY\DiazAN\DenefKN\GopakumarQS\ for some previous computations of various free field partition functions in related contexts.

As usual, we consider thermal AdS$_{d+1}$
\eqn\ca{ ds^2 = {1\over \cos^2 \rho}\left(d\rho^2 +dt^2 +\sin^2 \rho d\Omega_{d-1}^2 \right)}
with $t\cong t+\beta$, and a free scalar field,
\eqn\cb{ S = {1\over 2} \int\! d^{d+1}\sqrt{g} \phi( -\nabla^2+ m^2)\phi~.}
The mass is related to the scaling dimension as
\eqn\cc{ m^2 = \Delta(\Delta-d)~.}
The partition function is
\eqn\cd{ \ln Z = -{1\over 2} \Tr \ln ( -\nabla^2 +m^2).}
Noting that
\eqn\ce{ {d\over dm^2} \ln Z = -{1\over 2} \Tr {1\over -\nabla^2 +m^2} }
we can write
\eqn\cf{ \ln Z = {1\over 2} \int_{m^2}^\infty\! dm^2 \int\! d^3 x\sqrt{g} G_\Delta^\beta(x,x) }
where the propagator is
\eqn\ce{ G_\Delta^\beta(x,y) = \langle x|  {1\over  -\nabla^2  +m^2 }| y\rangle~. }
The integration in \cf\ is over thermal AdS.
The propagator must respect the thermal periodicity, which can be implemented by a sum over images.  If $G_\Delta(x,x')$ is the global AdS propagator then we can write
\eqn\cf{  G_\Delta^\beta(x,y) = \sum_{n=-\infty }^\infty G_\Delta(x,y_{n\beta})~.}
Here $y_{n\beta}$ denotes the bulk point related to $y=(t,\rho,\Omega_a)$ by $n$ thermal translations, $y_{n\beta} = (t+n\beta,\rho,  \Omega_a)$.    Plugging this into \cf, the divergent $n=0$ term can be cancelled by a cosmological constant counterterm, leaving
\eqn\cg{  \ln Z =  \sum_{n=1}^\infty \int_{m^2}^\infty\! dm^2 \int\! d^{d+1} x\sqrt{g} G_\Delta(x,x_{n\beta})~, }
where we used the symmetry under $n\rt -n$.
To proceed, we need a convenient form for the propagator that will allow us to carry out the integration over AdS.   The standard form of the propagator is
\eqn\ch{ G_\Delta(x,y)  = C_\Delta e^{-\Delta \sigma} F\left(\Delta,{d\over 2},\Delta+1-{d\over 2};e^{-2\sigma}\right)~,}
with
\eqn\ci{ C_\Delta = {\Gamma(\Delta) \over 2\pi^{d/2}\Gamma(\Delta+1-{d\over 2}) }~.   }
$F$ denotes the ${_2}F_1$ hypergeometric function and $\sigma=\sigma(x,y)$ is the geodesic distance between $x$ and $y$.  For $d=2$ the hypergeometric function greatly simplifies and it is straightforward to carry out the AdS integration.  This can in principle be extended to all even $d$ where it is possible to write the hypergeometric function in terms of elementary functions.  However, for odd $d$ this is not possible, and the required integrals are very challenging.    Another option is to use a heat kernel representation as in \GopakumarQS; however, this is again efficient only for $d$ even, since in odd $d$ the heat kernel only has an integral representation; e.g. \grig.

Instead, we will use a ``spectral" representation of the propagator \PenedonesUE.  This arises from decomposing the propagator into harmonic functions, writing the harmonic functions in the split representation as an integral over the AdS boundary, and then doing the boundary integrals.   This yields
\eqn\cj{ G_\Delta(u) = \pi^{-h} \int_{-i\infty}^{i\infty} \! {dc\over 2\pi i } {1\over \Gamma(c)\Gamma(-c)}{1\over (\Delta-h)^2-c^2} \int_0^\infty {dtd\tb \over t\tb} t^{h+c}\tb^{h-c} e^{-(t+\tb)^2-2ut\tb}~.}
The variable $u$ is related to the geodesic distance as
\eqn\ck{ u(x,y) = -1+\cosh \sigma(x,y)~.}
The parameter $h$ is related to the AdS$_{d+1}$ dimension as $ h= d/2$.
The advantage of this expression is that the dependence on AdS coordinates is simple, allowing the AdS bulk integral to be performed easily.   The equivalence of \ch\ and \cj\ can be established using the Mellin-Barnes representation of the hypergeometric function;  see \PenedonesUE.

We focus on a single term in the sum of \cg\ with fixed $n$.   The geodesic distance between the points $x$ and $x_{n\beta}$ is
\eqn\cm{ \cosh \sigma(x,x_{n\beta}) = {\cosh (n\beta) \over \cos^2 \rho}- \tan^2 \rho~.}
It will be convenient to change variables from $\rho$ to the coordinate $w$ defined as
\eqn\cn{\eqalign{  w&  \equiv  \cosh \sigma(x,x_{n\beta}) - \cosh(n\beta) \cr
&= {\cosh (n\beta) \over \cos^2 \rho}- \tan^2 \rho - \cosh(n\beta)~.}}
The AdS integration now takes the form
\eqn\co{ \int\! d^{d+1} x\sqrt{g} (\ldots) = {q^{nh} \over (1-q^n)^d} \int_0^{\beta}\!dt \int \! d \Omega_{d-1}  \int_0^\infty\! dw (2w)^{{d-2 \over 2}}(\ldots)~, }
where $q=e^{-\beta}$ as usual.   This gives
\eqn\cp{  \int\! d^{d+1} x\sqrt{g} e^{-2t\tb u(x,x_{n\beta}) }=   \pi^h {\beta q^{nd} \over (1-q^n)^d}  (t\tb)^{-h} e^{-2 \big(\cosh(n\beta)-1\big)t\tb}~,   }
where we have used $\int\! d\Omega_{d-1} = { 2\pi^h \over \Gamma(d/2)}$.  The $dtd\tb$ integration can be done using the formula
\eqn\cq{\eqalign{  \int_0^\infty {dtd\tb \over t\tb} t^{c}\tb^{-c}   e^{-t^2-\tb^2 -2\cosh (n\beta)   t\tb} & =  \Gamma(c)\Gamma(-c)\cosh(n\beta c)~,   }}
which can be derived by Taylor expanding $e^{-2\cosh (n\beta)   t\tb}$, doing the integrals term by term, and resumming.  Using these results we have
\eqn\crz{ \ln Z = 2\sum_{n=1}^\infty {\beta q^{nd} \over (1-q^n)^d} \int_\Delta^\infty \! d\Delta (\Delta-h)\int_{-i\infty}^{i\infty}{dc\over 2\pi i} {\cosh(n\beta c) \over (\Delta-h)^2-c^2}~.}
The remaining integrals are elementary, and we find
\eqn\cs{ \ln Z = \sum_{n=1}^\infty {1\over n} {q^{n\Delta} \over (1-q^n)^d}~,}
in agreement with \bfz.

\subsec{Higher spin fields}

We now comment on the generalization to symmetric, traceless, tensor fields.   The expression \cg\ is replaced by
\eqn\cgz{  \ln Z =  \sum_{n=1}^\infty \int_{m_J^2}^\infty\! dm_J^2 \int\! d^{d+1} x\sqrt{g} \Tr \left[ \Pi_{\Delta,J} (x,x_{n\beta})\right]~, }
where the trace refers to contraction of pairs of indices associated to the two distinct positions appearing in the spin-$J$ propagator $\Pi_{\Delta,J}$.    For a rank-$J$ tensor the scaling dimension $\Delta$ is related to the mass parameter appearing in the action as $ m_J^2=\Delta(\Delta-d)-J$.      The expected result is proportional to the scalar result, with the proportionality factor given by the number of spin states \bi,
\eqn\chz{ \ln Z = d_J  \sum_{n=1}^\infty {1\over n} {q^{n\Delta} \over (1-q^n)^d}~,}

To proceed we can again use the spectral representation.  The generalization of \cj\ is
\eqn\ci{  \Tr \left[ \Pi_{\Delta,J} (u)\right]  ={1\over  \pi^{d/2}} \int_{-i\infty}^{i\infty} \! {dc\over 2\pi i } f(c) \int_0^\infty {dtd\tb \over t\tb} t^{d/2+c}\tb^{d/2-c} e^{-(t+\tb)^2-2t\tb u } P_{\Delta,J}(u,t\tb)}
where we are writing
\eqn\cia{ f(c) = {1\over \Gamma(c)\Gamma(-c) [(\Delta-h)^2-c^2]}~.}
and $P_{\Delta,J}(u,t\tb)$ is a polynomial in $u$ and $t\tb$.   In appendix B we use this to evaluate the $J=1$ partition, obtaining \ch\ with $d_1=d$.

Note that if we consider the $n=1$ term in \ch\ we get the character,
\eqn\ciaa{ \chi_{\Delta,J} = \int_{m_J^2}^\infty\! dm_J^2 \int\! d^{d+1} x\sqrt{g} \Tr \left[ \Pi_{\Delta,J} (x,x_{\beta})\right]~, }
a result which will be used in the following.

\newsec{Quartic contact interactions: simple examples}

In this section we discuss some simple examples of  anomalous dimensions coming from bulk contact interactions.  We will consider a pair of scalar fields $(\phi_1,\phi_2)$ with quartic contact interactions involving some number of derivatives.  At zero bulk coupling this theory has spin-$J$ double trace operators of dimension  $\Delta_1+\Delta_2+2n+ J$.    The idea is to write the partition function, computed to first order in the interaction, in the form
\eqn\daa{ \ln Z = \ln Z\big|_{\lambda=0} + \sum_{J=0}^\infty \sum_{n=0}^\infty  {\p\chi_{\Delta,J} \over \p\Delta}\Big|_{\Delta=\Delta_1+\Delta_2+2n+J}\gamma(n,J)+ \ldots }
This identifies the coefficients $\gamma(n,J)$ as the double-trace anomalous dimensions.
To write the bulk result in this form we will use \ciaa.
%
%
Using $m_J^2 = \Delta(\Delta-d)-J$ we have
\eqn\dac{ {\p\chi_{\Delta,J} \over \p \Delta} = -(2\Delta-d)  \int\! d^{d+1}x\sqrt{g} \Tr\big[ \Pi_{\Delta,J} \big]~,}
so that the anomalous dimensions may be identified according to
\eqn\dad{  \ln Z = \ln Z\big|_{\lambda=0} - \sum_{J=0}^\infty \sum_{n=0}^\infty  \Big( (2\Delta-d) \int\! d^{d+1}x\sqrt{g} \Tr\big[ \Pi_{\Delta,J} \big] \Big) \Big|_{\Delta=\Delta_1+\Delta_2+2n+J}\gamma(n,J)+ \ldots }
This is quite useful, since it will be easy to massage the partition function into this form.

\subsec{Non-derivative interaction}

We consider the action
\eqn\da{ S= \int\! d^{d+1}x\sqrt{g} \left( {1\over 2}\phi_1(-\nabla^2+m_1^2  )\phi_1 +{1\over 2}\phi_2(-\nabla^2+m_1^2  )\phi_2 +\lambda (\phi_1 \phi_2)^2 \right)  ~.}
To first order in $\lambda$ the thermal partition function is
\eqn\db{ \ln Z = \ln Z \big|_{\lambda=0} - \lambda \int\! d^{d+1}x\sqrt{g} G_{\Delta_1}^\beta(x,x) G_{\Delta_2}^\beta(x,x)  + O(\lambda^2)~.}
The thermal  propagators are given as a sum over images, as in \cf.   We focus here on anomalous dimensions for double trace operators, which are obtained from the $n=\pm 1$ terms in the image sum.  Using the symmetry under $n\rt -n$, we retain just the $n=1$ terms and multiply the result by $4$ to get
\eqn\dc{ \ln Z = \ln Z \big|_{\lambda=0} - 4\lambda \int\! d^{d+1}x\sqrt{g} G_{\Delta_1}(x,x_\beta) G_{\Delta_2}(x,x_\beta)  +  \ldots~.}
where the $\ldots$ now include both terms with $n\geq 1$ and higher order in $\lambda$.

To proceed we note the following form of the scalar propagator
\eqn\dd{  G_\Delta(x,y) = C_\Delta \big(2u)^{-\Delta}  F\left(\Delta,\Delta-{d-1\over 2},2\Delta-d+1;-{2\over u} \right)}
related to \ch\ by a standard hypergeometric identity, and $u=-1+\cosh \sigma(x,y)$ as before.
Writing
\eqn\de{ G_{\Delta_1}(x,y) G_{\Delta_2}(x,y) = \langle 0| \phi_1(x) \phi_2(x) \phi_1(y)\phi_2(y)|0\rangle }
and inserting a complete set of two-particle states built on primaries of  dimension $\Delta_1+\Delta_2+2n$  it is clear that there exists an identity of the form
\eqn\de{ G_{\Delta_1}(x,y) G_{\Delta_2}(x,y)= \sum_{n=0}^\infty  a_n^{(0)} G_{\Delta_1+\Delta_2+2n}(x,y)~. }
The coefficients are readily determined by comparing terms in the $1/u$ expansion, and are given by \av\ with $s=0$.
%
%
Using this result in \dc\ gives
\eqn\dg{ \ln Z = \ln Z \big|_{\lambda=0} - 4\lambda \sum_{n=0}^\infty  a_n^{(0)}\int\! d^{d+1}x\sqrt{g} G_{\Delta_1+\Delta_2+2n}(x,x_\beta) +  \ldots~.}
%
%
%
%
%
Comparing to \dad, we identify the double-trace anomalous dimensions as
\eqn\dj{ \gamma(n,0) = {2 a_n^{(0)} \lambda \over \Delta_1+\Delta_2+2n -h}~.  }

\subsec{Two-derivative interaction}

Using integration by parts and the equations of motion we can take the two-derivative interaction to be
\eqn\dk{ S_{{\rm int}} = \lambda \int\! d^{d+1} \sqrt{g} \phi_1 \nabla^\mu \phi_1 \phi_2 \nabla_\mu \phi_2 }
The contribution to the partition function at first order in $\lambda$ is
\eqn\dl{ \ln Z =\ln Z \big|_{\lambda=0} -\lambda \int\!d^{d+1}x\sqrt{g} g^{\mu\nu} \nabla^{(1)}_\mu G^\beta_{\Delta_1}(x,x) \nabla^{(1)}_\nu G^\beta_{\Delta_2}(x,x)+ \ldots  }
where $  \nabla^{(1)}_\mu $ denotes differentiation with respect to the first argument of the propagator.    As in the last subsection, the thermal propagator is given by a sum over images, and we retain only the $n=\pm 1$ terms corresponding to double trace operators.  Using the identities
\eqn\dm{ G_\Delta(x,y)=G_\Delta(y,x)~,\quad G_\Delta(x,x_{-\beta})=G_\Delta(x_\beta,x)}
we have
\eqn\dn{ \nabla_\mu^{(1)} G_\Delta(x,x_{-\beta}) =  \nabla_\mu^{(2)} G_\Delta(x,x_{\beta}) \equiv  \nabla_\mu^{(2)} G_\Delta(x,y)\Big|_{y=x_{\beta}}~.  }
Using this we obtain
\eqn\doz{\eqalign{ \ln Z =\ln Z \big|_{\lambda=0} & -2\lambda \int\!d^{d+1}x\sqrt{g}  g^{\mu\nu}   \nabla_\mu^{(1)} G_{\Delta_1}(x,x_{\beta})   \nabla_\nu^{(1)} G_{\Delta_2}(x,x_{\beta}) \cr
&  -2\lambda \int\!d^{d+1}x\sqrt{g}  g^{\mu\nu}   \nabla_\mu^{(1)} G_{\Delta_1}(x,x_{\beta})   \nabla_\nu^{(2)} G_{\Delta_2}(x,x_{\beta})+ \ldots }}
We tackle the two terms in succession.

For the first term, if we apply $(\nabla^{(1)})^2$ to the identity \de, and use $(\nabla^{(1)})^2 G_\Delta(x,y) = \Delta(\Delta-d)G_\Delta(x,y)$ we have
\eqn\dpz{\eqalign{&  g^{\mu\nu}   \nabla_\mu^{(1)} G_{\Delta_1}(x,x_{\beta})   \nabla_\nu^{(1)} G_{\Delta_2}(x,x_{\beta})  =\sum_{n=0}^\infty c_n^{(0)}  G_{\Delta_1+\Delta_2+2n}(x,x_\beta)  }}
where we defined
\eqn\dq{ c_n^{(0)}  = {1\over 2} \Big[ (\Delta_1+\Delta_2+2n)(\Delta_1+\Delta_2+2n-d)-\Delta_1(\Delta_1-d)-\Delta_2(\Delta_2-d)  \Big] a_n^{(0)}~. }

Turning to the second term, we seek an identity of the form
\eqn\dr{\eqalign{   & \nabla_\mu^{(1)} G_{\Delta_1}(x,y)   \nabla_\nu^{(2)}  G_{\Delta_2}(x,y) \cr
 & \quad= \sum_{n=0}^\infty  a_n^{(1)}  \Pi_{\mu;\nu}(\Delta_1+\Delta_2+2n+1;x,y)+\sum_{n=0}^\infty b_n^{(0)}  \nabla_\mu^{(1)} \nabla_\nu^{(2)} G_{\Delta_1+\Delta_2+2n}(x,y)~.}}
Here $\Pi_{\mu;\nu}(\Delta;x,y)$ is the spin-1 propagator.  It obeys $(\nabla^2 - m_1^2) \Pi_{\mu;\nu}(\Delta;x,y)=0$   and $\nabla^\mu \Pi_{\mu;\nu}(\Delta;x,y) =0$  (for $x\neq y$)  with  $m_1^2 = \Delta(\Delta-d)-1$.    The existence of the identity \dr\ follows from the fact that the left hand side is a rank $(1,1)$ bitensor.  The general such bitensor can be written as a sum of scalar functions multiplying each of the two independent bitensors. The spin-1 propagator and the differentiated spin-0 propagator give us two linear combinations of these bitensors, and the spectrum of conformal dimensions is then chosen to match the expansion of the left hand side. In \dr\ we are anticipating that the coefficients multiplying the spin-1 propagators will turn out to be the $s=1$ coefficients defined in \av.

We will review spinning propagators in section 5, and here just note the following salient facts about the spin-1 propagator.   We can write
\eqn\ds{ \Pi_{\mu;\nu}(\Delta;x,y) = -{\p^2 u \over \p x^\mu \p y^\nu} g_0(\Delta;u)+{\p u \over \p x^\mu} {\p u \over \p y^\nu} g_1(\Delta;u)~. }
As in \ck\ $u$ is related to the geodesic distance as $u(x,y) =-1+\cosh \sigma(x,y)$.
The coefficient functions take the form
\eqn\dt{ g_0(\Delta;u) = G_\Delta(u) - h_1(\Delta;u)~,\quad g_1 = h_1'(\Delta;u)~,}
where $G_\Delta(u) $ is the usual scalar propagator, and $h_1(\Delta;u)$ is a function built out of $G_\Delta(u) $,
\eqn\dta{ h_1(\Delta;u) =- {(d-1) G_\Delta(u) +(1+u) G'_\Delta(u) \over (\Delta-1)(d-\Delta-1)}~.}
Going back to \dr\ and equating coefficients of the two tensor structures gives the pair of equations
\eqn\du{\eqalign{ &G'_{\Delta_1}(u) G'_{\Delta_2}(u)  =   \sum_{n=0}^\infty  a_n^{(1)}  h_1'(\Delta_1+\Delta_2+2n+1;u) +  \sum_{n=0}^\infty b_n^{(0)}  G''_{\Delta_1+\Delta_2+2n}(u) \cr
&0  = -\sum_{n=0}^\infty  a_n^{(1)}   \Big( G_{\Delta_1+\Delta_2+2n+1}(u) - h_1(\Delta_1+\Delta_2+2n+1;u)\Big) + \sum_{n=0}^\infty b_n^{(0)}   G'_{\Delta_1+\Delta_2+2n}(u)    }}
Differentiating the second equation and subtracting it from the first gives
\eqn\dv{  G'_{\Delta_1}(u) G'_{\Delta_2}(u)  = \sum_{n=0}^\infty  a_n^{(1)}   G'_{\Delta_1+\Delta_2+2n+1}(u)~,}
which is indeed the $s=1$ identity \at\  with coefficients defined in \av.
%
%
We then solve the first equation in \du\ to determine the $b_n^{(0)}$ as,
\eqn\dwa{ b_n^{(0)} = {1\over 8} \left[  1- \left( {\Delta_1(\Delta_1-d)-\Delta_2(\Delta_2-d) \over (\Delta_1+\Delta_2+2n)(\Delta_1+\Delta_2+2n-d)}\right)^2 \right] a_n^{(0)}~.}

Having determined all the coefficients we return to \dl\ and write the result in the form \bl,
\eqn\dx{\eqalign{ \ln Z = \ln Z\big|_{\lambda=0}  - \sum_{J=0}^1 \sum_{n=0}^\infty  \Big( (2\Delta-d) \int\! d^{d+1}x\sqrt{g} \Tr\big[ \Pi_{\Delta,J} \big] \Big) \Big|_{\Delta=\Delta_1+\Delta_2+2n+J}\gamma(n,J)+ \ldots }}
This is straightforward, and we read off the following anomalous dimensions:
\eqn\dy{\eqalign{ \gamma(n,1) & = {a_n^{(1)} \over \Delta_1+\Delta_2+2n+1-h}\lambda \cr
 \gamma(n,0) & = {c_n^{(0)} - (\Delta_1+\Delta_2+2n) (\Delta_1+\Delta_2+2n-d)b_n^{(0)} \over \Delta_1+\Delta_2+2n-h} \lambda~.     }}
To obtain this we used
\eqn\dz{ \int\! d^{d+1}x\sqrt{g}g^{\mu\nu}  \nabla_\mu^{(1)} \nabla_\nu^{(2)} G_\Delta(x,x_\beta) = -\Delta(\Delta-d)  \int d^{d+1}x\sqrt{g}G_\Delta(x,x_\beta)~,}
obtained by using integration by parts and the field equation.

%

\newsec{Spinning propagators}

As illustrated in our simple examples, our approach is based on taking a product of differentiated scalar propagators and expanding it in terms of spinning propagators and their derivatives.  For interaction vertices involving scalar fields, the relevant spinning propagators involve symmetric traceless tensors.  In this section we review these spinning propagators following \CostaKFA.  For earlier work see \AllenWD\DHokerBVE\NaqviVA.

\subsec{Embedding space}

It will be useful to work in embedding space, taking AdS$_{d+1}$ to be the hyperboloid $X\cdot X = \eta_{MN}X^M X^N =-1$ embedded in  $R^{d+1,1}$ with  line element $ds^2 = \eta_{MN}dx^M dx^N$ where
\eqn\ea{ \eta_{MN}X^M X^N = \sum_{a=1}^d (X^a)^2+ (X^{d+1})^2- (X^{d+2})^2~.}
The geodesic distance  (on the hyperboloid) between two points is
\eqn\eb{ \cosh \sigma(x,y) = -X\cdot Y~.}
The $u$ variable is then
\eqn\ec{ u(x,y)=-1+\cosh \sigma(x,y) = -1-X\cdot Y~.}
Global coordinates are defined by
\eqn\ed{\eqalign{ X^a & = \tan \rho \xh^a \cr
X^{d+1} & = {\sinh t\over \cos \rho}\cr
X^{d+2} & = {\cosh t\over \cos \rho} }}
with $\sum_{a=1}^d (\xh^a)^2=1$.  The corresponding metric is
\eqn\eez{ ds^2 = {1\over \cos^2 \rho}\left(d\rho^2 +dt^2 +\sin^2 \rho d\Omega_{d-1}^2 \right)~.}

We will be interested in symmetric traceless tensors. We start from a symmetric traceless embedding space tensor $T_{M_1, \ldots M_n}$ that has vanishing contraction with the normal vector to the hyperboloid, $X^{M_1} T_{M_1, \ldots M_n}=0$.   We then pull it back to the hyperboloid to obtain the AdS tensor,
\eqn\ef{ T_{\mu_1 \ldots \mu_n} = {\p X^{M_1} \over \p x^{\mu_1} } \ldots  {\p X^{M_n} \over \p x^{\mu_n} } {\p X^{M_1} \over \p x^{\mu_1} }T_{M_1, \ldots M_n} ~.}
Rather than display the indices, it is convenient to work with polynomials of polarization vectors $W_M$.     For traceless tensors we can use lightlike polarizaton vectors $W\cdot W =0$.  We can also impose $W\cdot X=0$, since we are assuming our tensors have no components normal to the hyperboloid.  Given the polynomial corresponding to $W^{M_1} \ldots W^{M_n} T_{M_1, \ldots M_n}$ we can extract a unique symmetric traceless tensor; see \CostaKFA\ for details.

\subsec{Spinning propagators}

In AdS$_{d+1}$ tensor language, the propagator for  a symmetric traceless tensor field obeys
\eqn\eg{\eqalign{ & (\nabla^2 -m_J^2) \Pi^{\Delta,J}_{\mu_1\ldots \mu_J;\nu_1\ldots \nu_J}(x,y) =0 \cr
&\nabla^{\mu_1}  \Pi^{\Delta,J}_{\mu_1\ldots \mu_J;\nu_1\ldots \nu_J}(x,y) =0 \cr
&  g^{\mu_1 \mu_2} \Pi^{\Delta,J}_{\mu_1\ldots \mu_J;\nu_1\ldots \nu_J}(x,y) =0~, }}
where $m_J^2 = \Delta(\Delta-d)-J$, and  \eg\ holds up to delta function terms on the right hand side for $x=y$.  Passing to embedding space, associated to the two points $X$ and $Y$ are two polarization vectors, obeying
\eqn\eh{ W_X \cdot W_X =  W_Y \cdot W_Y =  W_X \cdot X = W_Y \cdot Y = 0~.}
The general form of the propagator is governed by the fact that it is a rank $(J,J)$ bitensor.   The spin-$J$ propagator can be written in terms of $J+1$ scalar functions as
\eqn\ei{ \Pi_{\Delta,J}(X,Y)=\sum_{k=0}^J (W_{XY})^{J-k} \big( W_X \cdot \nabla_X W_Y \cdot \nabla_Y \big)^k f^{\Delta,J}_k(u)~.}
Here
\eqn\eia{  W_X \cdot \nabla_X = (W_X)^M {\p \over \p X^M}~,\quad W_{XY} = W_X\cdot W_Y~.}
A key fact is that the $k=0$ function is the scalar propagator,
\eqn\ej{ f^{\Delta,J}_0(u)=G_\Delta(u)~.}
The  $f^{\Delta,J}_k(u)$ for $k>0$  are determined iteratively in terms of $f^{\Delta,J}_0(u)$, in order to satisfy \eg.   Defining
\eqn\el{ h_k^{\Delta,J}(u) = (\p_u)^k f^{\Delta,J}_k(u)~,}
the relations are
\eqn\emz{  h_k^{\Delta,J} = c_k \Big( (d-2k+2J-1)\big( (d+J-2)h_{k-1}^{\Delta,J} +(1+u)\p_u h_{k-1}^{\Delta,J}\big) +(2-k+J)h_{k-2}^{\Delta,J}\Big)~,}
with
\eqn\en{ c_k =-{1+J-k \over k(d+2J-k-2)(\Delta+J-k-1)(d-\Delta+J-k-1)}~.  }
Using
\eqn\ek{ (W_X\cdot \nabla_X W_Y\cdot \nabla_Y)^n f(u) = \sum_{k=0}^n { (-1)^{n+k}\over   (n-k)!} \left( {n! \over k!}\right)^2  (W_{XY})^{n-k}(W_X \cdot Y W_Y\cdot X)^k (\p_u)^{n+k} f(u) }
we can re-express the propagator in the form
\eqn\eo{\eqalign{ \Pi_{\Delta,J}(X,Y) & = \sum_{k=0}^J (W_{XY})^{J-k}\big( W_X \cdot Y W_Y\cdot X \big)^k g_k^{\Delta,J}(u)  }}
%
with
\eqn\ep{ g_k^{\Delta,J}(u) = \sum_{n=k}^J {(-1)^{n+k}\over (n-k)!}  \left({n!\over k!}\right)^2  (\p_u)^{n+k} f_n^{\Delta,J} (u)~.  }

\newsec{General results}

\subsec{Interaction vertices}

In this section we discuss arbitrary bulk quartic contact interactions built out of a single scalar field.  We can use integration by parts and the free field equation to relate vertices.     The space of such vertices was described in \HeemskerkPN\ by associating them to flat space S-matrices built out of Mandelstam invariants. In particular, vertices with $2k$ derivatives are associated to monomials $s^a t^a u^c$ with $2k=4a+2c$ and where $0 \leq c \leq a$.   For example, there is a unique 4-derivative vertex, which we can take to be $\int (\nabla_\mu \phi \nabla^\mu \phi)^2$, corresponding to the monomial $st$.  A given vertex gives rise to anomalous dimensions for double trace operators with $J=0,2, \ldots, 2a$.    So at  $2k $-derivative order the  highest possible spin contribution is $J=k$, and we can take the corresponding vertex to be
\eqn\fa{ S_{\rm int} = \lambda \int\! d^{d+1}x\sqrt{g} \phi^2 (\nabla_{\mu_1} \ldots \nabla_{\mu_J} \phi)^2~.}

\subsec{Decomposition into spinning propagators}

Given some particular vertex, at first order in $\lambda$ the contribution to the partition function is obtained from the various Wick contractions among fields appearing in the vertex.  The resulting object to be integrated over thermal AdS is some index contraction of an object of type
\eqn\fb{  \nabla_{\mu_1} \ldots \nabla_{\mu_m} \nabla_{\nu_1}\ldots \nabla_{\nu_n} G_\Delta(x,y)   \nabla_{\mu_{m+1}} \ldots \nabla_{\mu_p} \nabla_{\nu_{n+1}}\ldots \nabla_{\nu_q} G_\Delta(x,y)~,}
where $y$ denotes a thermal translation of $x$: $y=x_\beta$.  In the above, we are using the convention that $\nabla_\mu$ acts on $x$, and $\nabla_\nu$ acts on $y$. As was illustrated in our simple examples, the strategy is to expand \fb\ in terms of spinning propagators and their derivatives; the coefficients in the expansion essentially yield the anomalous dimensions.

To facilitate this, note that we can always express an AdS tensor in terms of  a sum of symmetric traceless tensors combined with metric tensors.  Therefore, without loss of generality we can assume that \fb\ is symmetric and traceless in the $\mu$-type indices, and in the $\nu$-type indices.  Note that this implies that we have $q=p$ since we need to eventually contract each $\mu$ index with a $\nu$ index.  In the embedding space language of the last section, the product \fb\ then appears as
\eqn\fc{ (W_X\cdot  \nabla_X)^m (W_Y \cdot \nabla_Y)^n G_\Delta(X,Y) (W_X \cdot \nabla_X)^{p-m} (W_Y\cdot  \nabla_Y)^{p-n} G_\Delta(X,Y)~.}
The next step is to compute the functions $p_n$ appearing in the identity
\eqn\fd{\eqalign{ & (W_X\cdot  \nabla_X)^m (W_Y \cdot \nabla_Y)^n G_\Delta(X,Y) (W_X \cdot \nabla_X)^{p-m} (W_Y\cdot  \nabla_Y)^{p-n} G_\Delta(X,Y) \cr
& \quad =      \sum_{n=0}^p (W_{XY})^{p-n} (W_X\cdot \nabla_X W_Y\cdot \nabla_Y)^{n} p_n(u)~.       }}
To obtain equations that determine $p_n(u)$ we use \ek\ to express both sides of \fd\  in terms of products of $W_{XY}$, $W_X \cdot Y$ and $W_Y\cdot X$, and then equate coefficients.    Once the identity \fd\ is established, we use \ei\ to expand in terms of spinning propagators.
\eqn\fe{\eqalign{  &  \sum_{n=0}^p (W_{XY})^{p-n} (W_X\cdot \nabla_X W_Y\cdot \nabla_Y)^{n} p_n(u) \cr
& =  \sum_{n=0}^p \Big( C_n^{(p)} \Pi_{2\Delta+2n+p,p} + C_n^{(p-2)}(W_X\cdot \nabla_X W_Y \cdot \nabla_Y) \Pi_{2\Delta+2n+p-2,p-2}\cr
& \quad\quad\quad + \ldots + C_n^{(0)}(W_X\cdot \nabla_X W_Y \cdot \nabla_Y)^p \Pi_{2\Delta+2n,0}  \Big)    }}
for some constants $C_n^{(2s)}$.
This provides us with the decomposition of \fb\ into spinning propagators and derivatives thereof.   In the interaction vertex the indices are all contracted, and so all the derivatives either annihilate the propagators using the divergence free condition in \eg, or (possibly after integrating by parts) are Laplacians, which can be replaced by the corresponding  $m^2$ using \eg.   We are left with an expansion in terms of integrals of traced propagators, and as in our simple examples, the coefficients yield the anomalous dimensions.

\subsec{Highest spin contribution}

To illustrate the general procedure with an important example, in this section we work out the spin-$J$ anomalous dimensions induced by the vertex \fa.  The two distinct  Wick contractions  can be integrated by parts to the same form modulo terms that do not contribute at spin-$J$,
\foot{To get a maximal spin contribution the indices must be sequestered into two halves, with no index contractions occurring within one set.  This follows from the fact that a spin-$J$ two-particle state is created by a field bilinear with $J$ uncontracted indices acting on the vacuum.}
\eqn\ff{ \ln Z = \ln Z\big|_{\lambda =0} -4\lambda \int\! d^{d+1}x\sqrt{g} \nabla^{(1)}_{\mu_1} \ldots \nabla^{(1)}_{\mu_J}G^\beta_\Delta(x,x) \nabla_{(2)}^{\mu_1} \ldots \nabla_{(2)}^{\mu_J}G^\beta_\Delta(x,x) + \ldots }
the $\ldots$ now denoting terms at higher order in $\lambda$ and/or contribute only to spins $s<J$.  We replace the thermal propagators by a sum over images and keep just the $n=\pm 1$ terms, yielding
\eqn\fg{ \ln Z = \ln Z\big|_{\lambda =0} -16\lambda \int\! d^{d+1}x\sqrt{g} \nabla^{(1)}_{\mu_1} \ldots \nabla^{(1)}_{\mu_J}G_\Delta(x,x_\beta) \nabla_{(2)}^{\mu_1} \ldots \nabla_{(2)}^{\mu_J}G_\Delta(x,x_\beta) + \ldots }
Our task now is to expand the bilinear (using our convention that $(\mu,\nu)$ indices correspond to $(x,y)$ respectively)
\eqn\fh{ \nabla_{\mu_1} \ldots \nabla_{\mu_J}G_\Delta(x,y) \nabla_{\nu_1} \ldots \nabla_{\nu_J}G_\Delta(x,y) }
in terms of spinning propagators and their derivatives.   For present purposes we are just interested in the coefficient of the spin-$J$ propagator. In embedding space language  \fh\  corresponds to
\eqn\fiz{ (W_X \cdot \nabla_X)^J G_\Delta(u) (W_Y \cdot \nabla_Y)^J G_\Delta(u) =  (W_X \cdot Y W_Y \cdot X)^J  G^{(J)}_\Delta(u) G^{(J)}_\Delta(u)~,}
where we used that $u=-1-X\cdot Y$ and the notation \au.
Following our general strategy, the next step is to write
\eqn\fj{  (W_X \cdot Y W_Y \cdot X)^J G^{(J)}_\Delta(u) G^{(J)}_\Delta(u)= \sum_{n=0}^J (W_{XY})^{J-n} (W_X\cdot \nabla_X W_Y\cdot \nabla_Y)^{n} p_n(u)   ~.}
The system of equation determining $p_n(u)$ is found by using \ek\ on the right hand side and equating powers of $W_{XY}$.
As we will explain momentarily, we only need to work out $p_0(u)$, so it is convenient to take a linear combination of equations and their derivatives that isolates this function.   Let $E_q$ be the equation corresponding to the $(W_{XY})^{J-q}$ term.  It is then straightforward to verify that the linear combination $\sum_{q=0}^J q! (\p_u)^{J-q} E_q$ is
\eqn\fk{ J! G^{(J)}_\Delta(u) G^{(J)}_\Delta(u)= (\p_u)^J p_0(u)~.}
Using the identity
\eqn\fl{\eqalign{   G^{(s)}_{\Delta_1}(u)  G^{(s)}_{\Delta_2}(u)  & =   \sum_{n\geq 0} a^{(s)}_n   G^{(s)}_{\Delta_1+\Delta_2+2n+s}(u)  }}
with $a^{(s)}_n$ given in \av\
%
%
we have (with $\Delta_1=\Delta_2=\Delta$)
\eqn\fn{ p_0(u) = J! \sum_{n=0}^\infty a_n^{(J)} G_{2\Delta+2n+J}(u)~. }
With this result in hand we consider the expansion of \fj\ in terms of spinning propagators and their derivatives.  From \ei\ it is clear that a term proportional to $(W_{XY})^J$ can only come from the spin-$J$ propagator, since lower spin propagators will come with additional factors of $(W_X \cdot \nabla_X W_Y\cdot \nabla_Y)$ attached.  Therefore, our result for $p_0(u)$ immediately gives us the spin-$J$ contributions,
\eqn\fo{   (W_X \cdot \nabla_X)^J G_\Delta(u) (W_Y \cdot \nabla_Y)^J G_\Delta(u) = J! \sum_{n=0}^\infty a_n^{(J)} \Pi_{2\Delta+2n+J,J} + \ldots~.}
Using this result in \fg\ gives
\eqn\fp{ \ln Z = \ln Z\big|_{\lambda=0} -16J! \lambda \sum_{n=0}^\infty  a_n^{(J)} \int\! d^{d+1}x\sqrt{g} \Tr [  \Pi_{2\Delta+2n+J,J}] + \ldots~,}
and we then read off the spin-$J$ anomalous dimensions from \dad,
\eqn\fq{ \gamma(n,J) = {8J! a_n^{(J)}   \lambda\over 2\Delta +2n+J -h}~.}
We also note that if we considered replacing the single field $\phi$ with two distinct fields $\phi_1$ and $\phi_2$, then \fq\ holds if we use the general expression \av, replace $2\Delta \rightarrow \Delta_1 +\Delta_2$, and divide by a factor of two since there is now a single Wick contraction rather than two.

The result \fq\ was originally obtained in \HeemskerkPN\ by considering a four-point function in the Regge limit, which picks out the highest spin contribution.   Equation 5.44 of \HeemskerkPN\ agrees with \fq\ upon using the free field OPE coefficients found in \FitzpatrickDM.

\subsec{Another example}

We now consider the interaction
\eqn\fpa{ S_{\rm int} = \lambda \int\! d^{d+1}x\sqrt{g} \nabla_\mu \phi \nabla^\mu \phi (\nabla_{\mu_1} \ldots \nabla_{\mu_J} \phi)^2~.}
with $J$ even.
By working out the corresponding Mandelstam monomial we can see that this $2J+2$ derivative vertex has highest spin contribution given by spin-$J$. and we now work out the corresponding anomalous dimensions.  When considering the thermal diagram, with $y=x_\beta$ we will have  the contributing structure
\eqn\fpaa{\eqalign{ & \langle \nabla_\mu \phi(x) \nabla_{\mu_1} \ldots \nabla_{\mu_J}\phi(x) \nabla^\mu \phi(y) \nabla^{\mu_1} \ldots \nabla^{\mu_J}\phi(y)\rangle~.
  }}
We further form symmetric traceless combinations of each set of indices (this step does not affect the leading spin contribution). We think of inserting a complete of states in between the $x$ and $y$ operators.  The claim is that only spins up to $J$ contribute.  It might appear that spin-$(J+1)$ can contribute, but of course this cannot happen since for a single scalar there are no odd-spin two particle states.   We then open up the indices and try to establish a relation
\eqn\fpab{\eqalign{& \langle \nabla_\mu \phi(x) \nabla_{\mu_1} \ldots \nabla_{\mu_J}\phi(x) \nabla^\mu \phi(y) \nabla^{\mu_1} \ldots \nabla^{\mu_J}\phi(y)\rangle \cr
& \quad \quad\quad \quad\quad \quad\quad \quad = \sum_{n=0}^\infty \alpha_n \nabla_\mu \nabla_\nu [\Pi_{\Delta_{n,J},J}]_{\mu_1 \ldots \mu_J;\nu_1 \ldots \nu_J} +\ldots}}
where $\ldots$ denote lower spin contributions, and %
\eqn\faba{\Delta_{n,J} \equiv 2\Delta+2n+J~.}
On the left hand side there are two distinct Wick contractions.   In embedding space we then consider the equation
\eqn\fpac{\eqalign{ &  W_X \cdot \nabla_X(W_Y \cdot \nabla_Y)^J G_\Delta(u)  (W_X \cdot \nabla_X)^J W_Y \cdot \nabla_Y G_\Delta(u)\cr
& \quad +  W_X \cdot \nabla_X W_Y \cdot \nabla_YG_\Delta(u)(W_X \cdot \nabla_X W_Y \cdot \nabla_Y)^J G_\Delta(u) \cr
& \quad\quad = \sum_{n=0}^{J} (W_{XY})^{J-n} (W_X\cdot \nabla_X W_Y\cdot \nabla_Y)^{n+1} p_n(u)~. }}
As in our last example, knowledge of $p_0(u)$ will determine the spin$-J$ contribution.  If we use the identity
\eqn\fpad{\eqalign{  &  W_X \cdot \nabla_X (W_Y \cdot \nabla_Y)^J G_\Delta(u)  (W_X \cdot \nabla_X)^J W_Y \cdot \nabla_Y G_\Delta(u)\cr
& \quad +  W_X \cdot \nabla_X W_Y \cdot \nabla_YG_\Delta(u)(W_X \cdot \nabla_X W_Y \cdot \nabla_Y)^J G_\Delta(u) \cr
& ={1\over 2} W_X \cdot \nabla_X W_Y \cdot \nabla_Y\Big\{ \sum_{m,n=1}^{J-1}(-1)^{m+n} (W_X\cdot \nabla_X)^m (W_Y\cdot \nabla_Y)^n G_\Delta(u) \cr
 & \quad\quad\quad\quad\quad\quad\quad\quad\quad\quad\quad \times (W_X\cdot \nabla_X)^{J-m} (W_Y\cdot \nabla_Y)^{J-n} G_\Delta(u) \Big\}~,           }}
then the needed relation becomes
\eqn\fpae{\eqalign{& {1\over 2} \sum_{m,n=1}^{J-1}(-1)^{m+n} (W_X\cdot \nabla_X)^m (W_Y\cdot \nabla_Y)^n G_\Delta(u)  (W_X\cdot \nabla_X)^{J-m} (W_Y\cdot \nabla_Y)^{J-n} G_\Delta(u) \cr
& \quad = \sum_{n=0}^{J} (W_{XY})^{J-n} (W_X\cdot \nabla_X W_Y\cdot \nabla_Y)^{n} p_n(u) }}
We can isolate $p_0(u)$ just like we did in the previous example, forming the combination $\sum_{q=0}^J q! (\p_u)^{J-q}E_q$.   In this case we find
\eqn\fpaf{ {1\over 2} (J-1)^2 J! G_\Delta^{(J)}(u) G_\Delta^{(J)}(u)  = (\p_u)^J p_0(u)~.}
The solution is then the same as \fn\ up to an overall factor,
\eqn\fpag{ p_0(u) =  {1\over 2} (J-1)^2 J! \sum_{n=0}^\infty a_n^{(J)} G_{2\Delta+2n+J}(u)~. }
This determines the coefficients in \fpab\ as
\eqn\fpah{ \alpha_n =  {1\over 2} (J-1)^2 J!a_n^{(J)}~.}
We now insert the identity into the contribution to the partition function from \fpa.  We integrate once by parts, using that the spin$-J$ propagator has vanishing divergence and obeys $\nabla^2 \Pi_{\Delta_{n,J},J} = \big( \Delta_{n,J}(\Delta_{n,J}-d)-J\big)\Pi_{\Delta_{n,J},J}$.  This yields  (a factor of $4$ comes from the sum over images)
\eqn\fpai{ \ln Z = \ln Z\big|_{\lambda=0} -4(J-1)^2J! \lambda \sum_{n=0}^\infty   \big( \Delta_{n,J}(\Delta_{n,J}-d)-J\big) a_n^{(J)} \int\! d^{d+1}x\sqrt{g} \Tr [  \Pi_{2\Delta+2n+J,J}] + \ldots~,}
and we then read off the spin-$J$ anomalous dimensions from \dad,
\eqn\fpaj{ \tilde{\gamma}(n,J) =2(J-1)^2 J!  \big( \Delta_{n,J}(\Delta_{n,J}-d)-J\big) {a_n^{(J)}  \over 2\Delta +2n+J -h} \lambda~.}

As a check, this can be compared to the $d=2$ result computed in \HeemskerkPN\ from applying the bootstrap.  Our results yield the ratio between $\tilde{\gamma}(n,J)$ and the anomalous dimensions $\gamma(n,J)$ computed in the last section,
\eqn\fpak{\eqalign{ & {\tilde{\gamma}(n,J) \over \gamma(n,J)} = {1\over 2} (J-1)^2 \big(\Delta_{n,J}(\Delta_{n,J}-d)-J\big) \cr
& =   {1\over 2} (J-1)^2  \Big( J(2J+1)+(2\Delta+2n-1)(2\Delta+2n+2J-1)-(J+1)^2   \Big)  \quad (d=2)  }}
The bottom line was written to facilitate comparison to 4.16 and 4.19 of \HeemskerkPN.  The ratio given by those results in \HeemskerkPN\ agrees with \fpak\ up to overall normalization which is not fixed in \HeemskerkPN, except that the $(J+1)^2$ term in \fpak\ is absent.  The latter discrepancy is explained by the fact that in \HeemskerkPN\ they are discarding contributions that would come from a $(\nabla \phi)^4$ vertex, which would contribute an $n$ independent contribution to the ratio \fpak.  The point is that the $n$ dependent terms match as they should.

\newsec{ $(\nabla_\mu \phi \nabla^\mu \phi)^2$ interaction}

We present one more example in detail.
The four-derivative  interaction
\eqn\ga{S_{\rm int} = \lambda \int\! d^{d+1}x\sqrt{g} (\nabla_\mu \phi \nabla^\mu \phi)^2 }
yields anomalous dimensions for spin-$2$ and spin-$0$ double-trace primaries.  Since we can use integration by parts and the lowest order field equations to write $  (\nabla_\mu \phi \nabla^\mu \phi)^2 = \phi^2 (\nabla_{\mu}\nabla_\nu \phi)^2  $ plus terms with fewer derivatives, the computation of the spin-$2$ anomalous dimensions is a special case of section 6.   Extracting the spin-$0$ anomalous dimensions requires additional work.  For $d=2$ these were computed in \HeemskerkPN\FitzpatrickZM; here we compute these for arbitrary $d$ and verify agreement with previous results for $d=2$.

There are two distinct Wick contractions contributing to the partition function
\eqn\gb{\eqalign{  \ln Z_\lambda & =-\lambda  \int\! d^{d+1}x\sqrt{g} \Big( \nabla_\mu^{(1)}\nabla^\mu_{(2)} G^\beta_\Delta(x,x)\nabla_\nu^{(1)}\nabla^\nu_{(2)} G^\beta_\Delta(x,x)\cr
& \quad\quad\quad\quad\quad\quad\quad\quad +2 \nabla_\mu^{(1)}\nabla_\nu^{(2)} G^\beta_\Delta(x,x)\nabla^\mu_{(1)}\nabla^\nu_{(2)} G^\beta_\Delta(x,x) \Big)~.  }}
Summing over the $n=\pm 1$ thermal images and using the identities \dm-\dn, we have
\eqn\gc{ \ln Z_\lambda  =-4\lambda \int\! d^{d+1}x\sqrt{g}   K^{\mu_1 \mu_2;\nu_1 \nu_2}  \nabla_{\mu_1}^{(1)}\nabla_{\nu_1}^{(2)} G_\Delta(x,x_\beta )\nabla_{\mu_2}^{(1)}\nabla_{\nu_2}^{(2)} G_\Delta(x,x_\beta)    }
where
\eqn\gd{  K_{\mu_1 \mu_2;\nu_1 \nu_2}  = g_{\mu_1\nu_1}g_{\mu_2 \nu_2} + g_{\mu_1 \mu_2} g_{\nu_1 \nu_2}+g_{\mu_1 \nu_2}g_{\nu_1\mu_2}~. }
We separate out the traceless part by writing
\eqn\ge{  K_{\mu_1 \mu_2;\nu_1 \nu_2}  =  L_{\mu_1 \mu_2;\nu_1 \nu_2} + {d+3 \over d+1} g_{\mu_1 \mu_2 }g_{\nu_1 \nu_2} }
with the symmetric traceless tensor $ L_{\mu_1 \mu_2;\nu_1 \nu_2}$ defined as
\eqn\gf{ L_{\mu_1 \mu_2;\nu_1 \nu_2} =  g_{\mu_1\nu_1}g_{\mu_2 \nu_2} +g_{\mu_1 \nu_2}g_{\nu_1\mu_2} -{2\over d+1 }g_{\mu_1 \mu_2 }g_{\nu_1 \nu_2}~. }
There are two contributions to the partition function,
\eqn\gg{ \ln Z_\lambda = \ln Z_\lambda^{(1)} + \ln Z_\lambda^{(2)} }
where

\eqn\gg{\eqalign{  \ln Z^{(1)}_\lambda  &=-4\lambda \left({d+3 \over d+1}\right)  \int\! d^{d+1}x\sqrt{g} \nabla_{\mu}^{(1)}\nabla_{\nu}^{(2)} G_\Delta(x,x_\beta )\nabla^{\mu}_{(1)}\nabla^{\nu}_{(2)} G_\Delta(x,x_\beta) \cr
 \ln Z^{(2)}_\lambda  &=-4\lambda \int\! d^{d+1}x\sqrt{g}   L^{\mu_1 \mu_2;\nu_1 \nu_2}  \nabla_{\mu_1}^{(1)}\nabla_{\nu_1}^{(2)} G_\Delta(x,x_\beta )\nabla_{\mu_2}^{(1)}\nabla_{\nu_2}^{(2)} G_\Delta(x,x_\beta)~. }}
 $\ln Z_\lambda^{(1)}$ is easily dealt with using
\eqn\gh{\eqalign{&  \nabla_{\mu}^{(1)}\nabla_{\nu}^{(2)} G_\Delta(x,y )\nabla^{\mu}_{(1)}\nabla^{\nu}_{(2)} G_\Delta(x,y)\cr
&=  \left[{1\over 2}\nabla_x^2 -  \Delta(\Delta-d)  \right]\left[{1\over 2}\nabla_y^2 -  \Delta(\Delta-d)  \right]G_\Delta(x,y)G_\Delta(x,y)\cr
& = \left[{1\over 2}\nabla_x^2 -  \Delta(\Delta-d)  \right]\left[{1\over 2}\nabla_y^2 -  \Delta(\Delta-d)  \right]\sum_{n=0}^\infty a_n^{(0)} G_{2\Delta+2n}(x,y) \cr
& =    \sum_{n=0}^\infty\Big[2(\Delta+n)(\Delta+n-h) -  \Delta(\Delta-d)  \Big]^2 a_n^{(0)} G_{2\Delta+2n}(x,y)~,          }}
yielding
\eqn\gi{ \ln Z_\lambda^{(1)} =-4\lambda\left({d+3 \over d+1}\right)   \sum_{n=0}^\infty\Big[2(\Delta+n)(\Delta+n-h) -  \Delta(\Delta-d)  \Big]^2 a_n^{(0)} \int\! d^{d+1} x\sqrt{g} G_{2\Delta+2n}(x,x_\beta)~. }

Turning to $\ln Z^{(2)}_\lambda $ we now work out the coefficients in the expansion (recall our convention that $\mu/\nu$ indices refer to $x/y$)
\eqn\gj{\eqalign{&  \nabla_{\mu_1} \nabla_{\nu_1} G_\Delta(x,y )\nabla_{\mu_2}\nabla_{\nu_2} G_\Delta(x,y)  \cr
 & =  \sum_{n=0}^\infty \Big(d_n \nabla_{\mu_1 } \nabla_{\mu_2}\nabla_{\nu_1 }\nabla_{\nu_2 } G_{2\Delta+2n}(x,y) + 2a_n^{(2)} \Pi_{\mu_1\mu_2;\nu_1\nu_2}^{2\Delta+2n+2,2}(x,y) \Big)+{\rm traces}~.    }}
We have anticipated the fact, to be verified momentarily, that the spin-$2$ coefficients are $2a_n^{(2)}$.

In embedding space we consider
\eqn\gk{\eqalign{&  W_X \cdot \nabla_X W_Y \cdot \nabla_Y G_\Delta(u) W_X \cdot \nabla_X W_Y \cdot \nabla_Y G_\Delta(u)\cr
& \quad = \sum_{n=0}^\infty\Big(d_n (W_X \cdot \nabla_X W_Y \cdot \nabla_Y)^2 G_{2\Delta+2n}   +2a_n^{(2)}\Pi_{2\Delta+2n+2,2} \Big)~.  }}
Following our strategy in section 6 we first consider
\eqn\gkz{ W_X \cdot \nabla_X W_Y \cdot \nabla_Y G_\Delta(u) W_X \cdot \nabla_X W_Y \cdot \nabla_Y G_\Delta(u)=      \sum_{n=0}^2 (W_{XY})^{2-n} (W_X\cdot \nabla_X W_Y\cdot \nabla_Y)^{n} p_n(u)~.    }
Applying \ek\ the coefficient functions are determined from
\eqn\gl{\eqalign{  p_0'' &= 2 G^{(2)}_\Delta G^{(2)}_\Delta  \cr
p_1'''
& = 4G^{(2)}_\Delta G^{(2)}_\Delta - ( G^{(1)}_\Delta G^{(1)}_\Delta)''  \cr
p_2'''' & = G^{(2)}_\Delta G^{(2)}_\Delta~.             }}
From the first equation we deduce
\eqn\gm{ p_0 =2 \sum_{n=0}^\infty a_n^{(2)} G_{2\Delta+2n+2}~. }
As in our previous examples, knowledge of $p_0$ fixes the highest spin contribution, here spin-$2$, and we confirm the spin-$2$ coefficients in \gj.

We now turn to the computation of $d_n$, for which there are various way to proceed.  One option is take the divergence of \gj\ which projects out the spin-$2$ terms.   We instead work in the $(W_{XY}, W_X\cdot Y W_Y \cdot X)$ basis and use the explicit form of the spin-$2$ propagator
\eqn\gn{\eqalign{ \Pi_{\Delta,2}(X,Y) & = \sum_{k=0}^2 (W_{XY})^{2-k}\big( W_X \cdot Y W_Y\cdot X \big)^k g_k^{\Delta,2}(u) ~. }}
%
The $(W_{XY})^2$ term in the equation \gk\ is
\eqn\go{\eqalign{ G^{(1)}_\Delta G^{(1)}_\Delta = 2\sum_{n=0}^\infty \Big( d_n G^{(2)}_{2\Delta+2n} +a_n^{(2)} g_0^{2\Delta+2n+2,2}   \Big) ~.          }}
Using the known expression for $ g_0^{2\Delta+2n+2,2}$ we solve this equation for $d_n$, obtaining
\eqn\gp{ d_n = \left( {(2h-1)n^2+(2h-1)(2\Delta-h)n+h\Delta(2\Delta-2h-1)\over 2h(2\Delta+2n+1)(2\Delta+2n-1-2h)} \right)^2 a_n^{(0)}~.  }
So, we have now determined the expansion \gj.

Returning to \gg\  we can use  integration by parts to write (under an integral sign
\eqn\gq{\eqalign{& L^{\mu_1\mu_2;\nu_1 \nu_2} \nabla_{\mu_1} \nabla_{\mu_2 }\nabla_{\nu_1} \nabla_{\nu_2 } G_\Delta(x,x_\beta) \cr
&\quad = 2 \Big( \Delta(\Delta-d)-d -{1 \over d+1} \Delta(\Delta-d)\Big)\Delta(\Delta-d)G_\Delta(x,x_\beta)\cr
&\quad  =  {2d\over d+1}\Delta (\Delta+1) (\Delta-d-1) (\Delta-d) G_\Delta(x,x_\beta)~. }}
We therefore have
\eqn\gr{\eqalign{ \ln Z_\lambda^{(2)} & = - 8\lambda \sum_{n=0}^\infty \Big( {d\over d+1}\Delta_n (\Delta_n+1) (\Delta_n-d-1) (\Delta_n-d)d_n \int\! d^{d+1}x\sqrt{g} G_{\Delta_n}(x,x_\beta) \cr
&\quad\quad\quad\quad\quad\quad  + 2a_n^{(2)}\int\! d^{d+1}x\sqrt{g}  \Tr[ \Pi_{\Delta_n+2,2}(x,x_\beta) ]     \Big)         }}
where
\eqn\gs{\Delta_n =2\Delta+2n~.}
Putting our results together, we can now read off the anomalous dimensions from \dad,
\eqn\gt{\eqalign{ \gamma(n,2) & = {8a_n^{(2)}  \over 2\Delta+2n +2 -h} \lambda \cr
\gamma(n,0) & = 2\left({d+3 \over d+1}\right)   {\big[2(\Delta+n)(\Delta+n-h) -  \Delta(\Delta-d)  \big]^2 \over 2\Delta+2n-h } a_n^{(0)}\lambda   \cr
& \quad + {4d\over d+1} { \Delta_n (\Delta_n+1) (\Delta_n-d-1) (\Delta_n-d) \over 2\Delta+2n-h } d_n \lambda  }}
Upon setting $d=2$, one can verify that this matches A.7 of \FitzpatrickZM, which in turn matches D.1 of \HeemskerkPN,  after taking into account that the latter authors throw away terms that would come from a no-derivative $\phi^4$ interaction.

\newsec{Discussion}
 
The purpose of this work was to develop an efficient approach to computing thermal AdS partition functions of weakly coupled scalar fields, both for its own sake and for extracting anomalous dimensions of double trace operators, as is relevant for the AdS/CFT correspondence.  We found that this provides a strikingly simple way of extracting anomalous dimensions induced by contact interactions, and in particular we were able to easily generalize known results to arbitrary spacetime dimension.  In our approach, no explicit AdS integrations need be performed, as these are all absorbed into the definition of the characters in terms of which the computation is expressed.  This simplification is analogous to that provided by the use of geodesic Witten diagrams in the computation of boundary correlation functions \Hijanozsa.   We worked out various illustrative examples in which we could make contact with previous results, but it should be clear that it is straightforward  to handle any scalar contact interaction, and we outlined the general procedure for doing so.

There are numerous natural directions in which to extend these results.  One is to  replace our scalar fields by fields with spin.    The same strategy will apply, with the new ingredient being that one needs to expand the product of two spinning propagators in terms of other spinning propagators.    Other  obvious directions to pursue are to include exchange interactions and higher loop effects. One would again like to organize the computation so as to avoid having to perform difficult AdS integrals; this will require the use of propagator identities that go beyond those implemented in this work.

\vskip.3cm
\noindent
{\bf Acknowledgements}
\vskip.2cm
We wish to thank David Meltzer, Eric Perlmutter, David Simmons-Duffin for discussions. We thank River Snively for discussions and collaboration during the early stages of this work. The research of PK is supported in part by the National Science Foundation under research grant PHY-16-19926. SM is grateful to the Alexander S. Onassis Foundation for its support. AS is supported by the College of Arts and Sciences of the University of Kentucky.

\appendix{A}{Scalar partition function with angular potentials}

In this appendix we discuss the computation of scalar  characters and partition function functions in the presence of angular potentials.

\subsec{CFT character}

Given the SO(d,2) symmetry group of CFT$_d$, we can take the Cartan generators to be the dilatation operator $H$ and the $2r$ Cartan generators of SO(d), where $d=2r$ or $d=2r+1$.   We write the latter generators as
\eqn\Cartan{
H_i = M_{2i-1,2i}~, \; \; \; i=1,...r  \; .
}
Given
\eqn\commrel{
[M_{a b}, P_c] =i (\delta_{a c} P_b   -  \delta_{b c}  P_a)  \; ,
}
we have that $P^\pm_j = P_{2j-1} \pm i P_{2j}$ obey
\eqn\za{[H,P^\pm_j]=P^\pm_j~,\quad  [H_i, P^\pm_j] = \pm  \delta_{ij} P^\pm_j~.}
The character is
\eqn\zb{ \chi_{\Delta,R}(\beta,\mu_i) = \Tr \left[ e^{-\beta H - i\sum_j \mu_j H_j}\right]~.}
We here restrict attention to scalar primaries, with $R$ being the singlet representation of SO(d);  for general results see \DolanWY.   Acting on the primary state with any string of $P^\pm_i$ we compute
\eqn\zc{\chi_{\Delta,0}(\beta,\mu_i)  = \left\{  \matrix{  {q^{\Delta}\over \prod_{j=1}^{r}(1-q y_j)(1-q /y_j)}  & d=2r \cr  \cr  {q^{\Delta}\over (1-q) \prod_{j=1}^{r}(1-q y_j)(1-q /y_j)}& d=2r+1  }   \right. }
with
\eqn\zd{ q=e^{-\beta}~,\quad y_j =e^{i\mu_j}~.}

\subsec{Free Partition Function in AdS}

The introduction of non-zero angular potentials is easily incorporated into the previous computation in section 3. We first consider the case of $d$ even and write $d=2r$.  In embedding space a thermal translation is now described as
\eqn\zl{\eqalign{ X^1 \pm X^{2r+2} & \rt X_\beta ^1 \pm X_\beta^{2r+2} = e^{\pm \beta} \big(X^1 \pm X^{2r+2}  \big)\cr
X^{2j} \pm i X^{2j+1} & \rt X_\beta^{2j} \pm i X_\beta^{2j+1} = e^{\pm i \mu_j } \big(  X_\beta^{2j} \pm i X_\beta^{2j+1} \big)~,\quad   j=1,2, \ldots r~.  }}
It is convenient to use coordinates adapted to these identifications,
\eqn\zla{\eqalign{ X^{2r+2}\pm X^1 & = \sqrt{1+ r_1^2+ \ldots + r_r^2} e^{\pm t}\cr
X^{2j} \pm i X^{2j+1} & = r_j e^{\pm i \phi_j}~.}}
The half-chordal distance between a point and its thermal image is
\eqn\zg{
u (x,x_{\beta})=-1 + \cosh \beta- \; r_1^2 (\cos \mu_1 - \cosh \beta) -...- \; r_r^2 (\cos \mu_r - \cosh \beta).
}
The AdS integral in \cp\ becomes
\eqn\adsint{\eqalign{
& \int d^{d+1}x \sqrt{g}  e^{-2u t\bar{t}} \cr
&\quad = \int dt \; \prod_{i}^{r} \int r_i dr_i \; \int d\phi_i e^{2t\bar{t} (1-\cosh \beta +r_1^2 (\cos \mu_1 - \cosh \beta) +...+ \; r_r^2 (\cos \mu_r - \cosh \beta) )}  \cr
&\quad= \beta (2 \pi)^r   e^{2 t\bar{t} (1-\cosh \beta)   }   \prod_{i=1}^r \bigg(  \int_0^{\infty} r_i dr_i \; e^{2 t\bar{t}  \; r_i^2  (\cos \mu_i - \cosh \beta)  }  \bigg) \cr
&\quad= \beta (2 \pi)^r   e^{2 t\bar{t} (1-\cosh \beta)   }   \prod_{i=1}^r \bigg( {1\over 2 t\bar{t}  \;  2( \cosh \beta - \cos \mu_i ) }    \bigg)  \; \cr
&\quad = \pi^h{ \beta q^d \over \prod_{i=1}^{d/2} (1-qy_i)(1-q/y_i)  } (t\tb)^{-h} e^{-2(\cosh\beta -1)t\tb}~.
}}
All the other integrals proceed in exactly the same way as in eq. \cp-\crz, yielding
\eqn\zh{
\ln Z_1 =   {q^{\Delta}  \over   \prod_{i=1}^{d/2}((1-qy_i)(1-q/y_i))  }       \; .
}
We wrote $Z_1$ since we are only considering the single winding contribution.
Similarly, for $d=2r+1$, we get
\eqn\oddcftpart{
\ln Z_1 ={q^{\Delta}\over (1-q) \prod_{i=1}^{r} (1-qy_i)(1-q/y_i)}
}
These results are in agreement with \zc.

\appendix{B}{Free spin-1 partition function}

The free spin-1 propagator was given in \ds,
\eqn\za{ (\Pi_\Delta)_{\mu;\nu}(x,y) = -{\p^2 u \over \p x^\mu \p y^\nu} g_0(\Delta;u)+{\p u \over \p x^\mu} {\p u \over \p y^\nu} g_1(\Delta;u)~. }
We compute its trace using the relations
\eqn\zb{\eqalign{ g^{\mu\nu} {\p^2 u \over \p x^\mu \p y^\nu} &=-(d-1)-(q+q^{-1})+u  \cr  g^{\mu\nu} {\p u \over \p x^\mu} {\p u \over \p y^\nu} &= (u-q-q^{-1})u }}
where we are using the fact that the metric is the same at the points $x$ and $y$ since they are related by a translation in $t$; we are working in global coordinates.   The functions $g_0$ and $g_1$ are given in \dt,\dta.  They are expressed in terms of $G_\Delta(u)$ and first and second derivatives thereof, with each term given by a degree one  polynomial $u$.   Starting from the representation \cj\ for $G_\Delta(u)$, and noting that each $u$ derivative just brings down a factor of $-2t\tb$, we arrive at the expression \ci,
\eqn\zc{  \Tr \left[ \Pi_{\Delta} (u)\right]  ={1\over  \pi^{d/2}} \int_{-i\infty}^{i\infty} \! {dc\over 2\pi i } f(c) \int_0^\infty {dtd\tb \over t\tb} t^{d/2+c}\tb^{d/2-c} e^{-(t+\tb)^2-2t\tb u } P_{\Delta}(u,t\tb)}
where $ P_{\Delta}(u,t\tb)$ is a degree $3$ polynomial in $u$  and a degree two polynomial in $t\tb$, whose explicit form is not particularly illuminating.  We now evaluate the partition function from \cgz,
\eqn\zd{  \ln Z =  \sum_{n=1}^\infty \int_{\Delta}^\infty\! d\Delta (2\Delta-d) \int\! d^3 x\sqrt{g} \Tr \left[ \Pi_{\Delta} (x,x_{n\beta})\right]~. }
The remaining steps are  straightforward and not particularly instructive to display in detail.  We first carry out the AdS integrals, followed by the $(t,\tb)$ integrals, and finally evaluate the $c$ integral by evaluating residues.   The last step involves one subtlety, which is that there are poles in the right half plane at $c=\Delta-h$ and also at $c=1$.   The former pole yields the desired partition function, which makes it clear that we should choose the integration contour to run to the right of the $c=1$ pole, a fact which we have not attempted to justify from first principles.   Taking this into account, it is straightforward to arrive at the expected result
\eqn\ze{ \ln Z = \sum_{n=1}^\infty {1\over n} {d q^{n\Delta} \over (1-q^n)^d}~.}
This same strategy can be applied to higher spins as well, though we expect the details to be more involved.

\appendix{C}{Anomalous dimensions from four-point Witten diagrams}

\lref\KarateevOML{
  D.~Karateev, P.~Kravchuk and D.~Simmons-Duffin,
  ``Harmonic Analysis and Mean Field Theory,''
JHEP {\bf 1910}, 217 (2019).
[arXiv:1809.05111 [hep-th]].
}

\lref\SleightFPC{
  C.~Sleight and M.~Taronna,
  ``Spinning Witten Diagrams,''
JHEP {\bf 1706}, 100 (2017).
[arXiv:1702.08619 [hep-th]].
}

\lref\MeltzerNBS{
  D.~Meltzer, E.~Perlmutter, and A.~Sivaramakrishnan,
  ``Unitarity Methods in AdS/ CFT'',
JHEP {\bf 2003}, 061 (2020).
[arXiv:1912.09521 [hep-th]].
}

Progress in the conformal bootstrap program has led to powerful methods for computing anomalous dimensions. In this section, we review an efficient bootstrap-based method that is the natural sibling of the partition function approach taken in this paper. We decompose the four-point Witten diagram of interest into conformal blocks, and then extract the anomalous dimensions from the coefficients in this expansion. This section will discuss this method for derivative contact tree diagrams developed in \CostaKFA, \BekaertTVA, and closely follows the review in \MeltzerNBS, to which we refer the reader for further details and subtleties omitted here. We focus on the contact diagram with vertex $\lambda \phi_1 ( \nabla_{\mu_1} \ldots \nabla_{\mu_J} \phi_2) \phi_3 ( \nabla^{\mu_1} \ldots \nabla^{\mu_J}  \phi_4)$.

Working in embedding space\foot{We denote integrals over AdS as $\int_{AdS} dY$ , and over the boundary as $\int_{\partial AdS} dX $ .}   (reviewed in Section 5), the (symmetric part of) the derivative contact diagram we are interested in is
\eqn\ASa{\eqalign{
{\cal A}^{\phi^4}_J (x_i) = {\lambda \over \left( J! \left( {d-1 \over 2}\right)_J \right)^2}\int_{AdS} d Y  & K_{\Delta_1}(Y, X_1 ) (K  \cdot \nabla)^J K_{\Delta_2}(Y, X_2 )
\cr
&
K_{\Delta_3}(Y, X_3 ) (K  \cdot \nabla)^J  K_{\Delta_4}(Y,X_4 ),
}}
where $K_{\Delta} \equiv K_{\Delta,0}$ is the spin-0 bulk-to-boundary propagator \foot{Explicitly, $K_{\Delta,J}(Y,P; W,Z) = { \cal C}_{\Delta,J}
{\left( (-2P\cdot Y)(W \cdot Z)+2(W \cdot P) (Z \cdot Y))\right)^J \over (-2 P \cdot Y)^{\Delta+J}}$, with normalization ${ \cal C}_{\Delta,J} = {(\Delta-1+J)\Gamma(\Delta) \over 2\pi^{d/2} (\Delta-1)\Gamma(\Delta+1-d/2)}$. We will often suppress the last two arguments of $K_{\Delta,J}$ for brevity.}, and we abuse notation to denote the projector onto AdS by $K$ (see e.g \CostaKFA, \PenedonesUE\ for further details).   We will first use the spinning completeness relation
\eqn\SpinningCompleteness{
\sum_{l=0}^J \int d\nu c_{J,J-l}(\nu)((W_1 \cdot \nabla_1)(W_2 \cdot \nabla_2))^{J-l} \Omega_{\nu,l}(Y_1,Y_2,W_1,W_2) = \delta(Y_1,Y_2)(W_{12})^J,
}
where the spectral function is  \CostaKFA
\eqn\ASb{
c_{J,l}(\nu) = {2^l(J-l+1)_l(h+J-l-1/2)_l \over l!(2h+2J-2l-1)_l (h+J-l-i \nu)_l(h+J-l+i\nu)_l},
}
where we refer the reader to \CostaKFA, \PenedonesUE\ for a definition and properties of the harmonic  function $\Omega_{\nu,l}$  not immediately needed here.
Using the completeness relation, the contact diagram becomes
\eqn\ASc{\eqalign{
 {\cal A}_{J}^{\phi^4}(x_i)& =
{\lambda \over \left( J! \left( {d-1 \over 2}\right)_J \right)^2}
\sum_{l=0}^J  \int d\nu c_{J,J-l}(\nu)
\int_{AdS}  dY_A dY_B\cr
 & \quad \times K_{\Delta_1}(Y_A,X_1) (K_A \cdot \nabla_{A})^J K_{\Delta_2}(Y_A,X_2)
 ((W_1 \cdot \nabla_1)(W_2 \cdot \nabla_2))^{J-l}\cr
 & \quad \times  \Omega_{\nu,l}(Y_A,Y_B,W_1,W_2)
K_{\Delta_3}(Y_B,X_3) (K_B \cdot \nabla_{B})^J  K_{\Delta_4}(Y_B,X_4).
}}
The spins of the operators in the block expansion are at this point determined, as we will see shortly. Now, we use the split representation
\eqn\SpinningSplitRep{
\Omega_{\Delta_\nu, J}(Y_1,Y_2; W_1,W_2) = {\nu^2 \over \pi J!(h-1)_J} \int_{\partial AdS} dP K_{\Delta_\nu,J}(Y_1,P;W_1,D_Z) K_{\widetilde{\Delta}_\nu,J}(Y_2,P;W_2,Z)
}
to write the contact diagram as products of AdS three-point functions. We use the subscript $\nu$ to indicate dimensions lying on the principal series $\Delta_\nu = {d \over 2}+ i \nu$ with real spectral parameter $\nu$. We denote the dimension of the shadow operator as $\widetilde{\Delta} = d-\Delta$. The diagram becomes
\eqn\ASd{\eqalign{
&{\cal A}_{J}^{\phi^4}(x_i) =
{\lambda \over \left( J! \left( {d-1 \over 2}\right)_J \right)^2}
\sum_{l=0}^J \int
{  c_{J,J-l}(\nu) \nu^2 d\nu \over \pi l!(h-1)_l} \int_{\partial AdS} dP
\cr
&
\int_{AdS}  dY_A K_{\Delta_1}(Y_A,X_1) (K_A \cdot \nabla_{A})^J K_{\Delta_2}(Y_A,X_2)
(W_1 \cdot \nabla_1)^{J-l}K_{\Delta_\nu,J}(Y_A,P;W_1,D_Z)
\cr
&
\int_{AdS}  dY_B
K_{\Delta_3}(Y_B,X_3) (K_B \cdot \nabla_{B})^J  K_{\Delta_4}(Y_B,X_4) (W_2 \cdot \nabla_2)^{J-l}
 K_{\widetilde{\Delta}_\nu,J}(Y_B,P;W_2,Z)
.
}}
The three-point integrals are known:
\eqn\ASe{\eqalign{
{1 \over J! \left( {d-1 \over 2}\right)_J }
\int_{AdS}  dY
&
K_{\Delta_1}(Y,X_1) (K_A \cdot \nabla_{A})^J K_{\Delta_2}(Y,X_2)
(W_3 \cdot \nabla_3)^{J-l}K_{\Delta_3,l}(Y,X_3;W_3,Z)
\cr
&= b(\Delta_1,\Delta_2,\Delta_3,l,J) \langle {\cal O}_1(X_1) {\cal O}_2(X_2) {\cal O}_3(X_3, Z) \rangle,
 }}
where $b(\Delta_1,\Delta_2,\Delta_3,l,J)$ was computed in \CostaKFA. The spinning three-point structure, defined as the three-point function without the OPE coefficient, is
\eqn\ASf{
\langle {\cal O}_1(P_1) {\cal O}_2(P_2) {\cal O}_3(P_3, Z) \rangle =
{\left(
(Z \cdot P_1)P_{23}- (Z \cdot P_2) P_{13}
\right)^J
\over
P_{13}^{ {\Delta_1+\Delta_3-\Delta_2+J_3 \over 2}}
P_{23}^{ {\Delta_2+\Delta_3-\Delta_1+J_3 \over 2}}
P_{12}^{ {\Delta_1+\Delta_2-\Delta_3+J_3 \over 2}}
}
. }
We then have
\eqn\ASg{\eqalign{
{\cal A}_{J}^{\phi^4}(x_i) &= \sum_{l=0}^J \int
 {  c_{J,J-l}(\nu) \nu^2 d\nu \over \pi l!(h-1)_l}
  b(\Delta_1,\Delta_2,\Delta_\nu,J,l)b(\Delta_3,\Delta_4,\widetilde{\Delta}_\nu,J,l)
\cr
&~~~~~ \times \int_{\partial AdS} dP \langle {\cal O}_1(X_1) {\cal O}_2(X_2) {\cal O}_\nu (P, D_Z) \rangle
 \langle \widetilde{{\cal O}}_\nu (P, Z) {\cal O}_3(X_3) {\cal O}_4(X_4) \rangle
.
 }}
The above integral of two three-point structures defines the conformal partial wave $\Psi^{1234}_{\Delta,J}(x_i)$, which is related to the conformal block as\foot{The normalization of the conformal block we use is
$
K^{\Delta_1,\Delta_2}_{\Delta,J} =
\left(- {1 \over 2} \right)^J
 {\pi^{d/2} \Gamma\left( \Delta - {d \over 2}  \right) \Gamma(\Delta+J-1) \Gamma\left(  {\widetilde{\Delta}+J+\Delta_1-\Delta_2 \over 2}   \right) \Gamma\left(  {\widetilde{\Delta}+J+\Delta_2-\Delta_1 \over 2}   \right)
\over
\Gamma(\Delta-1)\Gamma \left( \widetilde{\Delta}+J \right)
\Gamma\left(  {\Delta+J+\Delta_1-\Delta_2 \over 2}   \right) \Gamma\left(  {\Delta+J+\Delta_2-\Delta_1 \over 2}   \right) }.
$
}
\eqn\ASi{
\Psi^{1234}_{\Delta_\nu,l}(x_i)
=K^{\Delta_3,\Delta_4}_{\widetilde{\Delta}_\nu,l} g^{1234}_{\Delta_\nu,l}(x_i) + K^{\Delta_1,\Delta_2}_{\Delta_\nu,l } g^{1234}_{\widetilde{\Delta}_\nu,l}(x_i).
 }
The contact diagram becomes
\eqn\ASj{
{\cal A}_{J}^{\phi^4}(x_i) = \sum_{l=0}^J \int
 {  c_{J,J-l}(\nu) \nu^2 d\nu \over \pi l!(h-1)_l}
 b(\Delta_1,\Delta_2,\Delta_\nu,J,l)b(\Delta_3,\Delta_4,\widetilde{\Delta}_\nu,J,l) \Psi^{1234}_{\Delta_\nu,l}(x_i) .
 }
We can close the $\nu$ contour in the lower half plane for $g^{1234}_{\Delta_\nu,l}(x_i)$ and the upper half for $g^{1234}_{\widetilde{\Delta}_\nu,l}(x_i)$ to obtain the conformal block decomposition, but we leave this step implicit for now. The poles in the $b$-factors determine which operators appear.
To proceed further, it is helpful to focus on a simple example, the maximal spin case $l=J$. The $b$-factors are
\eqn\SpinningBfactor{\eqalign{
&b_{\Delta_1,\Delta_2, \Delta,J,J} = {\cal C}_{\Delta_1,0} {\cal C}_{\Delta_2,0} {\cal C}_{\Delta,J}\cr
 &\quad \times
{\pi^{d/2}
\Gamma\left( {\Delta_1+\Delta_2-\widetilde{\Delta}+J \over 2}\right)
\Gamma\left( {\Delta_1+\Delta_2-\Delta+J \over 2}\right)
\Gamma\left( {\Delta_2+\Delta-\Delta_1+J \over 2}\right)
\Gamma\left( {\Delta+\Delta_1-\Delta_2+J \over 2}\right)
\over
2^{1-J} \Gamma(\Delta_1)\Gamma(\Delta_2)\Gamma(\Delta_3+J)}.
 }}
The poles in the integrand that contribute are at double trace locations $\Delta_{\nu} = \Delta_1+\Delta_2 + 2n+J$ and $ \Delta_3+\Delta_4 + 2n+J$, and their shadows\foot{The integrand is shadow symmetric, so we need only analyse the poles in the lower-half $\nu$ plane. See \MeltzerNBS\ for discussion of spurious poles in this context.}.
Anomalous dimensions arise as double-poles in spectral space; that is, when the poles corresponding to  $[{\cal O}_1 {\cal O}_2]_{n,J}$ and $ [{\cal O}_3 {\cal O}_4]_{n,J}$ exchange coincide. For simplicity, we will take $\Delta_3=\Delta_1$ and $\Delta_4=\Delta_2$. As first studied in \HeemskerkPN, the term in the block expansion that contains the anomalous dimensions is of the form $ T_s(x_i) \sum_{n,l} p^{(0)}_{n,l}\gamma_{n,l}^{(1)} \left({1 \over 2} \partial_n \right) g^{1234}_{\Delta_{n,l},l} (x_i) $, where $T_s (x_i)$ is a purely kinematic prefactor that will not be important here and $p^{(0)}_{n,l}$ are the squared mean field theory coefficients \HeemskerkPN  \FitzpatrickDM  \KarateevOML,
\eqn\ASk{\eqalign{
& p^{(0)}_{n,l} = \cr
 & {(-1)^n(\Delta_1-h+1)_n(\Delta_2-h+1)_n (\Delta_1)_{l+n}(\Delta_2)_{l+n} \over l!n!(l+h)_n(\Delta_1+\Delta_2+n-2h+1)_n(\Delta_1+\Delta_2+2n+l-1)_l (\Delta_1+\Delta_2+n+l-h)_n}.
 }}
Putting everything together, the anomalous dimensions are therefore\foot{Note that for the maximal spin case, $c_{J,0}(\nu) = 1$.}
\eqn\ASl{
\gamma_{n,J} = - {  4(\Delta_{n,J}-h)^2    K^{\Delta_1, \Delta_2}_{\widetilde{\Delta}_{n,J},J} \over J!(h-1)_J p^{(0)}_{n,J}}
{\rm Coeff}^2_{\Delta_\nu = \Delta_{n,l}}
~
\left [
b(\Delta_1,\Delta_2,\Delta_\nu,J,J)b(\Delta_1,\Delta_2,\widetilde{\Delta}_\nu,J,J)
\right]
,
 }
where Coeff$^2_{\Delta_\nu = \Delta_{n,l}}$ indicates the coefficient of the double pole. In this case, for non-negative integer $n$,
\eqn\xyz{ {\rm Coeff}^2_{n=0,1,2,\ldots }  ~ \left[ \Gamma^2(-n) \right] = {1 \over (n!)^2}.}
This result matches \ay\ when setting the operators identical, up to a factor independent of $n,\Delta$. For submaximal spins $(l \leq J)$ we need to include the trace contributions as well, which can also be computed as we have described using \CostaKFA.
To summarize, this approach required the completeness relation \SpinningCompleteness, the split representation \SpinningSplitRep, and three-point integrals \SpinningBfactor, and did not require the explicit blocks or solving crossing.\foot{At least, for the contact diagram. For the exchange diagram and certain loop diagrams \Liujhs, \MeltzerNBS, we will need to expand blocks in the crossed channel, which requires use of the explicit blocks or the $6j$ symbol of the Euclidean conformal group.} Once these identities are assembled, the block decomposition follows automatically, and the anomalous dimensions can be easily read off. Compared to the partition function approach, this is an indirect method of obtaining anomalous dimensions. However, much of the necessary computation has already been carried out, and conformal symmetry can be used to greatly simply the structure.
It would be interesting to derive identities like \at, where two propagators can be expanded in a basis of single propagators, using a similar approach. By embedding this identity in a four-point Witten diagram, and then equating the double discontinuities of the resultant bubble and tree diagrams (computed in \MeltzerNBS), one can derive the zero-derivative coefficients $a_n^{(0)}$. We leave a similar investigation of the derivative relations to future work.

\listrefs
\end

\newsec{Conventions and basic formulas}

Here I temporarily collect some formulas that will eventually find their way into the other sections.

We work with Euclidean AdS$_{d+1}$ with $L_{\rm AdS}=1$.     We often use $h=d/2$. The action for a free scalar field is
\eqn\aa{ S ={1\over 2} \int\! d^{d+1}\sqrt{g} \left( (\p \phi)^2 + m^2 \phi^2 \right)~.}
$\phi$ is dual to a CFT operator of dimension $\Delta$ where
\eqn\ab{ m^2 = \Delta(\Delta-d)~.}
The scalar propagator obeys
\eqn\ac{ \big(\nabla^2 -\Delta(\Delta-d)\big)G_\Delta(x,y) = -{1\over \sqrt{g} }\delta^{d+1}(x,y)~.}
It is given by
\eqn\ad{  G_\Delta(x,y) = C_\Delta \big(2u)^{-\Delta}  F\left(\Delta,\Delta-{d-1\over 2},2\Delta-d+1;-{2\over u} \right)}
with
\eqn\ae{ C_\Delta = {\Gamma(\Delta) \over 2\pi^{d/2}\Gamma(\Delta+1-{d\over 2}) }   }
$F$ denotes the ${_2}F_1$ hypergeometric function.  $u$ is related to the geodesic distance $\sigma(x,y)$  between the two points as
\eqn\af{ u(x,y) = -1+ \cosh \sigma(x,y)~.}
Another form for the propagator is
\eqn\ag{ G_\Delta(x,y)  = C_\Delta e^{-\Delta \sigma} F\left(\Delta,{d\over 2},\Delta+1-{d\over 2};e^{-2\sigma}\right)~.}

The spectral form of the bulk-bulk scalar propagator is
\eqn\al{ G_\Delta(u) = \pi^{-h} \int_{-i\infty}^{i\infty} \! {dc\over 2\pi i } {1\over \Gamma(c)\Gamma(-c)}{1\over (\Delta-h)^2-c^2} \int_0^\infty {dtd\tb \over t\tb} t^{h+c}\tb^{h-c} e^{-(t+\tb)^2-2ut\tb}~.}
(Here we follow the $u$  convention of the Spinning paper, as in \af.)

Writing
\eqn\ala{  G^{(s)}_{\Delta}(u) \equiv {d^s \over du^s} G_\Delta(u)}
the equation
\eqn\am{\eqalign{   G^{(s)}_{\Delta_1}(u)  G^{(s)}_{\Delta_2}(u)  & =   \sum_{n\geq 0} a^{(s)}_n   G^{(s)}_{\Delta_1+\Delta_2+2n+s}(u)  }}
is solved by
\eqn\an{\eqalign{  a^{(s)}_n  &= { (-2)^s (h+s)_n \over 2\pi^h n!}{(\Delta_1+\Delta_2+2n+2s)_{1-h-s} (\Delta_1+\Delta_2-2h+n+1)_{n}\over (\Delta_1+n+s)_{1-h-s}(\Delta_2+n+s)_{1-h-s} (\Delta_1+\Delta_2-h+n+s)_n}  }}
These coefficients can be rewritten as
\eqn\ao{\eqalign{  a^{(s)}_n  &= { (-2)^s (h+s)_n \over 2\pi^h n!}{ \Gamma(\Delta_1+n+s)\Gamma(\Delta_2+n+s)\Gamma(\Delta_1+\Delta_2+n+s-h)\over \Gamma(\Delta_1+n-h+1)\Gamma(\Delta_2+n-h+1)\Gamma(\Delta_1+\Delta_2+2n+2s)   } \cr
& \quad \times  (\Delta_1+\Delta_2+n-2h+1)_n (\Delta_1+\Delta_2+2n+s-h) }}

We will often be working in embedding space, with AdS$_{d+1}$ defined as the hyperboloid $X\cdot X=-1$. Here
\eqn\aga{ X\cdot X = \sum_{a=1}^d (X^a)+ (X^{d+1})^2- (X^{d+2})^2~.}
The relation to the geodesic distance is
\eqn\ah{ \cosh \sigma(x,y) = -X\cdot Y~.}
Note then that
\eqn\ai{ u(x,y) = -1-X\cdot Y~.}
Global coordinates are defined by
\eqn\aj{\eqalign{ X^a & = \tan \rho \xh^a \cr
X^{d+1} & = {\sinh t\over \cos \rho}\cr
X^{d+2} & = {\cosh t\over \cos \rho} }}
with $\sum_{a=1}^d (\xh^a)^2=1$.  The corresponding metric is
\eqn\ak{ ds^2 = {1\over \cos^2 \rho}\left(d\rho^2 +dt^2 +\sin^2 \rho d\Omega_{d-1}^2 \right)~.}
The volume of the unit $(d-1)$-sphere is
\eqn\akb{\Omega_{d-1} = {2\pi^{d\over 2} \over \Gamma\left(d\over 2\right) }}
From \ai, for two points at the same radial and angular location,
\eqn\aka{ u(x,x')  = {2\sinh^2 \left( {t-t' \over 2}\right) \over \cos^2 \rho}~.}
We denote the bulk-bulk propagator for a symmetric traceless rank-$J$ tensor as $\Pi_{\Delta,J}(x,y)$.   It obeys
\eqn\am{\eqalign{ & (\nabla^2 -m^2) \Pi^{\Delta,J}_{\mu_1\ldots \mu_J;\nu_1\ldots \nu_J}(x,y) =0 \cr
&\nabla^{\mu_1}  \Pi^{\Delta,J}_{\mu_1\ldots \mu_J;\nu_1\ldots \nu_J}(x,y) =0 \cr
&  g^{\mu_1 \nu_1} \Pi^{\Delta,J}_{\mu_1\ldots \mu_J;\nu_1\ldots \nu_J}(x,y) =0~, }}
where $m^2 = \Delta(\Delta-d)-J$, and the \am\ holds up to delta function terms on the right hand side for $x=y$.   Given two bulk points $x$ and $y$ our convention that $\mu$ type indices refer to $x$, and $\nu$ type indices to $y$.     We will work in embedding space.   To handle traceless symmetric tensors we introduce $W_X$ and $W_Y$, obeying
\eqn\an{ W_X \cdot W_X =  W_Y \cdot W_Y =  W_X \cdot X = W_Y \cdot Y = 0~.}
A polynomial in the $(W_X,W_Y)$ then defines a unique symmetric, traceless tensor on AdS. We will use the notation $W_{XY} = W_X \cdot W_Y$.   The propagator has the embedding space form
\eqn\ao{ \Pi_{\Delta,J}(X,Y)=\sum_{k=0}^J (W_{XY})^{J-k} \big( W_X \cdot \nabla_X W_Y \cdot \nabla_Y \big)^k f^{\Delta,J}_k(u)~.}
A key fact is that the $k=0$ function is the scalar propagator,
\eqn\ap{ f^{\Delta,J}_0(u)=G_\Delta(u)~.}
The other $f^{\Delta,J}_k(u)$ are determined iteratively in terms of $f^{\Delta,J}_0(u)$.  A useful relation is
\eqn\ekz{ (W_X\cdot \nabla_X W_Y\cdot \nabla_Y)^n f(u) = \sum_{k=0}^n { (-1)^{n+k}\over   (n-k)!} \left( {n! \over k!}\right)^2  (W_{XY})^{n-k}(W_X \cdot Y W_Y\cdot X)^k (\p_u)^{n+k} f(u) }
\eqn\gbfz{\eqalign{  \nabla_{\mu_1 }\nabla_{\mu_2} u &= (1+u)g_{\mu_1\mu_2} \cr
\nabla^\mu u \nabla_\mu u &= u(2+u)\cr
\nabla^\mu \nabla_\mu u &= (d+1)(1+u) \cr
\nabla_\nu \nabla_\mu u \nabla^\mu u & =(1+u)\nabla_\nu u  }}
\eqn\yxba{\eqalign{  \nabla_{\mu_1 }\nabla_{\mu_2} u &= (1+u)g_{\mu_1\mu_2} \cr
 \nabla_{\mu_1 }\nabla_{\mu_2}\nabla_\nu  u &= g_{\mu_1\mu_2} \nabla_\nu u\cr
\nabla^\mu u \nabla_\mu u &= u(2+u)\cr
\nabla^\mu \nabla_\mu u &= (d+1)(1+u) \cr
\nabla_\mu \nabla^\nu u \nabla_\nu u & =(1+u)\nabla_\mu u \cr
 \nabla_{\mu_1} \nabla^\nu u \nabla_{\mu_2} \nabla_\nu u & =\nabla_{\mu_1} u\nabla_{\mu_2 }u -u g_{\mu_1\mu_2}  }}
\eqn\yulaa{ \nabla^2 \nabla_\mu \phi = \nabla_\mu \nabla^2 \phi -d \nabla_\mu \phi}
For the last one see {\tt check 3 deriv ident.mw}